\documentclass[onecolumn, prb, reprint, amsfonts, amsmath, amssymb, superscriptaddress, longbibliography, floatfix]{revtex4-2}
\usepackage[T1]{fontenc}
\usepackage{graphicx}
\usepackage[dvipsnames]{xcolor}
\usepackage[colorlinks,linkcolor=Blue,citecolor=Blue]{hyperref}
\usepackage{braket}
\usepackage{wasysym}
\usepackage{MnSymbol}
\usepackage{bbm}
\usepackage{mathrsfs}

\usepackage[capitalise]{cleveref}

\crefname{section}{Sec.}{Secs.}
\Crefname{section}{Section}{Sections}

\crefname{appendix}{Appendix}{Appendices}
\Crefname{appendix}{Appendix}{Appendices}

\newcommand{\ci}{\mathbbm{i}}

\newcommand{\ham}{H}
\newcommand{\kin}{\mathcal{K}}
\newcommand{\pot}{\mathcal{V}}

\newcommand{\VAB}{{V_{AB}}}
\newcommand{\VS}{{V_{S}}}

\newcommand{\newcomment}[1]{}

\usepackage{tcolorbox}
\tcbuselibrary{theorems,skins}

\newcommand{\tilesymbol}[1]{\vcenter{\hbox{\includegraphics[height = 0.9\baselineskip]{#1}}}}
\newcommand{\tileA}{ \tilesymbol{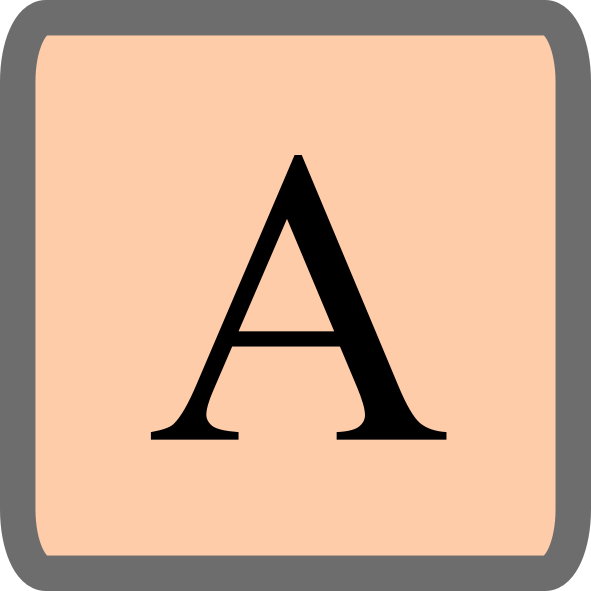}}
\newcommand{\tileB}{ \tilesymbol{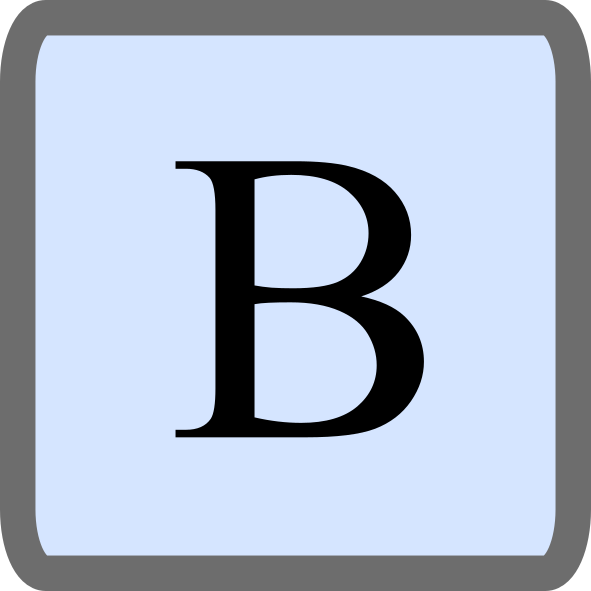}}
\newcommand{\tileS}{ \tilesymbol{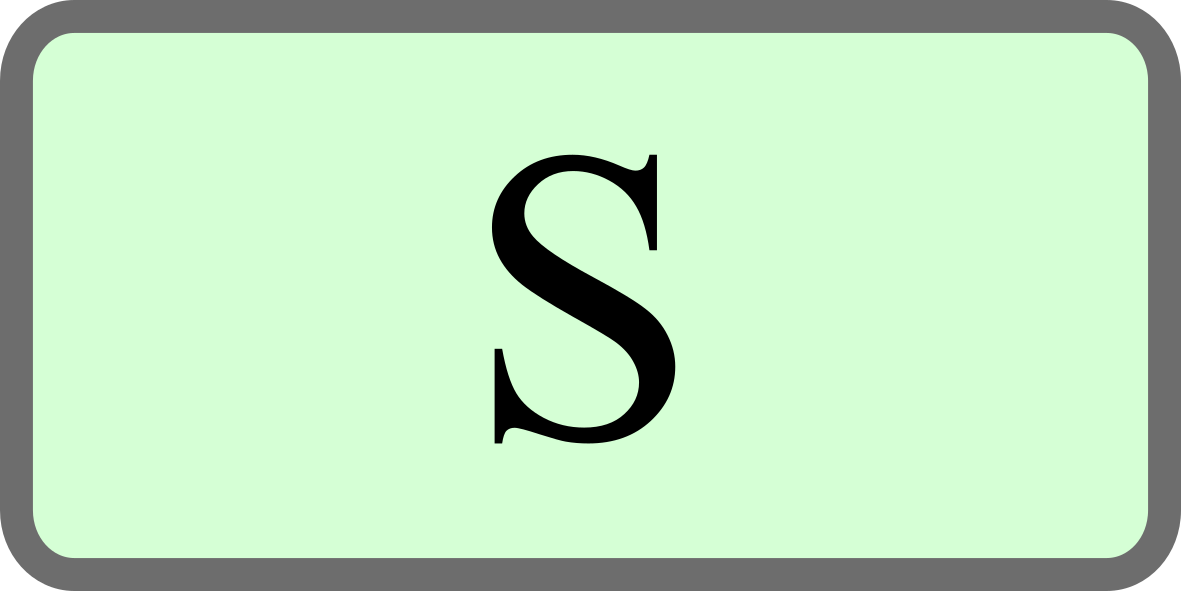}}
\newcommand{\tileX}{ \tilesymbol{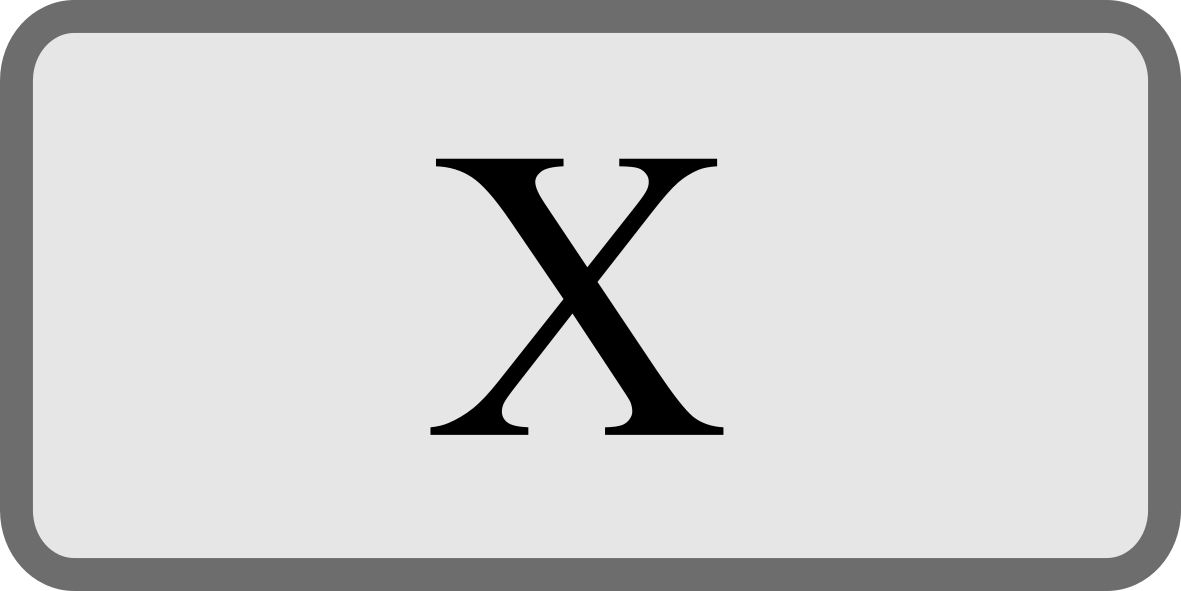}}
\newcommand{\tileY}{ \tilesymbol{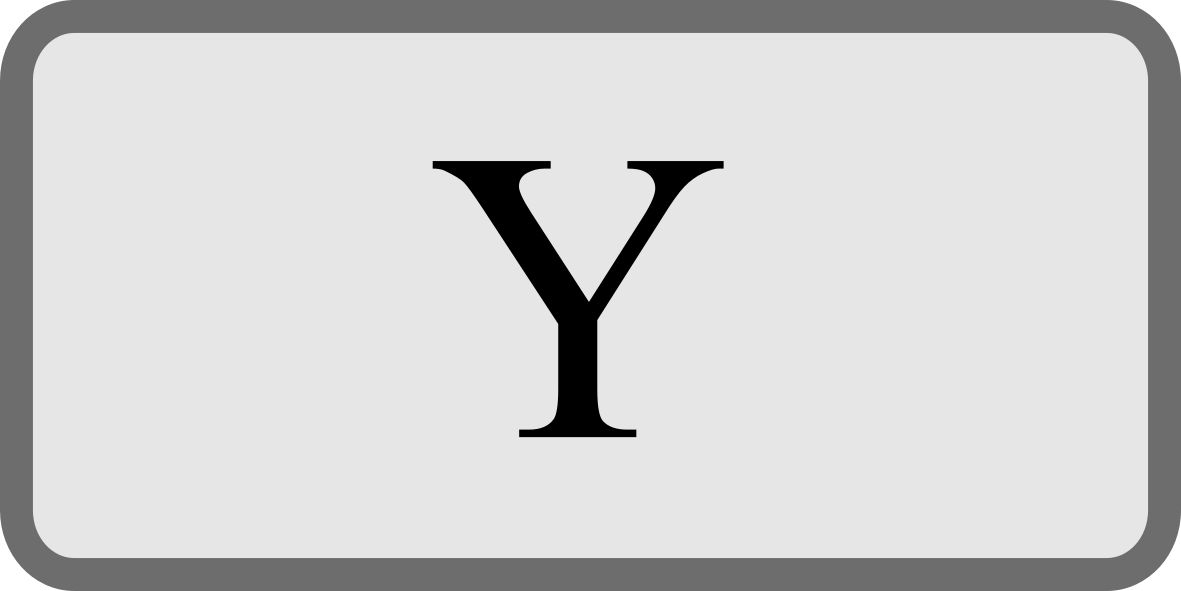}}
\newcommand{\tileAS}{ \tilesymbol{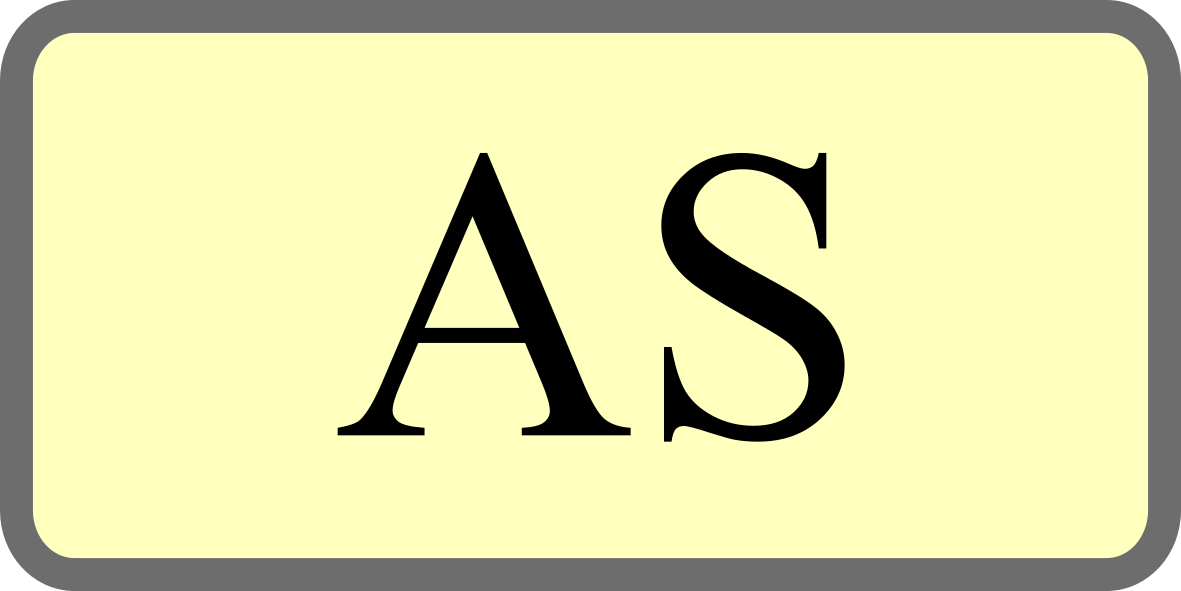}}

\newcommand{\gaugesymbol}[1]{\vcenter{\hbox{\includegraphics[height = 1.5em]{#1}}}}

\newcommand{\stringtriangleupflat}{\gaugesymbol{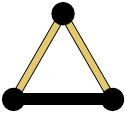}}
\newcommand{\stringtriangledownflat}{\gaugesymbol{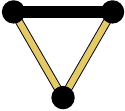}}
\newcommand{\stringtriangleupbent}{\gaugesymbol{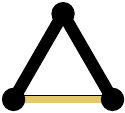}}
\newcommand{\stringtriangledownbent}{\gaugesymbol{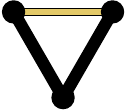}}
\newcommand{\stringsquareup}{\gaugesymbol{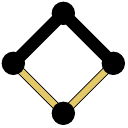}}
\newcommand{\stringsquaredown}{\gaugesymbol{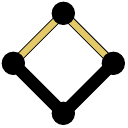}}
\newcommand{\stringtriangleupflatdimer}{\gaugesymbol{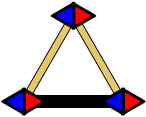}}
\newcommand{\stringtriangledownflatdimer}{\gaugesymbol{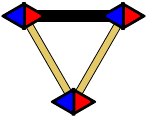}}
\newcommand{\stringtriangleupbentdimer}{\gaugesymbol{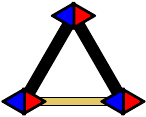}}
\newcommand{\stringtriangledownbentdimer}{\gaugesymbol{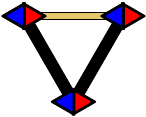}}
\newcommand{\stringsquareupdimer}{\gaugesymbol{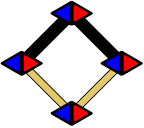}}
\newcommand{\stringsquaredowndimer}{\gaugesymbol{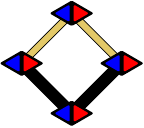}}

\newcommand{\dimersquarehorz}{\gaugesymbol{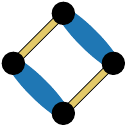}}
\newcommand{\dimersquarevert}{\gaugesymbol{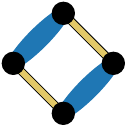}}
\newcommand{\dimerhexagonup}{\gaugesymbol{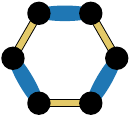}}
\newcommand{\dimerhexagondown}{\gaugesymbol{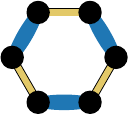}}

\newcommand{\GSD}{\textrm{GSD}}
\newcommand{\Appendix}[1]{\section{#1}}

\newcommand{\XXZ}[2]{\mathbf{XXZ}~[{#1},{#2}]}
\newcommand{\SpinOne}[2]{\mathbf{SpinOne}~[{#1},{#2}]}
\newcommand{\TileChain}[2]{\mathbf{TileChain}~[{#1},{#2}]}

\newcommand{\sqQVM}{{$\mathrm{QVM}_{\hspace{0.5 pt} \square}$}}
\newcommand{\trQVM}{{$\mathrm{QVM}_{\bigtriangleup}$}}
\newcommand{\sqQDM}{{$\mathrm{QDM}_{\hspace{0.5 pt} \square}$}}
\newcommand{\hxQDM}{{$\mathrm{QDM}_{\hspace{0.5 pt} \hexagon}$}}

\newcommand{\guided}[1]{}

\begin{document}

\title{A web of exact mappings from RK models to spin chains}

\author{Gurkirat Singh}
\affiliation{Center for Condensed Matter Theory, Indian Institute of Science, 560012, Bangalore}
\affiliation{Department of Physics, Massachusetts Institute of Technology, Cambridge, Massachusetts 02139, USA}

\author{Inti Sodemann Villadiego}
\email[sodemann@itp.uni-leipzig.de]{}
\affiliation{Institut f{\"u}r Theoretische Physik, Universit{\"a}t Leipzig, Br{\"u}derstra{\ss}e 16, 04103, Leipzig, Germany}

\date{\today}

\begin{abstract}
    We study Rokhsar-Kivelson (RK) dimer and spin ice models realizing $U(1)$-lattice gauge theories in a wide class of quasi-one-dimensional settings, which define a setup for the study of few quantum strings (closed electric field lines)  interacting with themselves and each other. We discover a large collection of mappings of these models onto three quantum chains: the spin-1/2 XXZ chain, a spin-1 chain, and a kinetically constrained fermion chain whose configurations are best described in terms of tilings of a rectangular strip. We show that the twist of boundary conditions in the chains maps onto the transverse momentum of the electric field string, and their Drude weight to the inverse of the string mass per unit length.  We numerically determine the phase diagrams for these spin chains, employing DMRG simulations and find global similarities but also many interesting new features in comparison to the full 2D problems. For example, the spin-1 chain we obtain features a continuous family of degenerate ground states at its RK point analogous to a Bloch sphere, but without an underlying microscopic global $SU(2)$ symmetry. We also argue for the existence of a (stable) Landau-forbidden gapless critical point away from the RK point in one of the models we study using bosonization and numerics. This is surprising given that the full 2D problem is generically gapped away from the RK point. The same model also displays extensively many local conserved quantities which fragment the Hilbert space, arising as a consequence of destructive resonances between the electric field lines.  Our findings highlight spin-chain mappings as a potent technique for the exploration of unusual dynamics, exotic criticality, and low-energy physics in lattice gauge theories.
\end{abstract}

\maketitle

    \section*{Introduction}
    Interactions in quantum many body systems often restrict the set of allowed configurations by imposing large energetic barriers to certain states. This can manifest as the presence of an extensive number of local constraints, such as in quantum spin ice \cite{QSI_Bramwell,QSI_Gingras}, where at low energies the spins are constrained to obey a 2-in-2-out rule. In many cases, such constraints resemble those found in lattice gauge theories \cite{IntrotoLGT_Kogut,QSFT_Wen,QSFT_Fradkin}, and can consequently realize unusual, fractionalized excitations and exotic quantum orders \cite{FractionalizationLecture_Laughlin,RVB_Dimer_Moessner,DefinedTO_Wen,QSFT_Fradkin}.
    
    In addition to their potential realization in frustrated magnets \cite{FisherBalentsPhotons,PyrochloreQSL}, there have been considerable recent advances in realizing constrained models in quantum simulators (such as Rydberg atoms \cite{Rydberg_forQI,Rydberg_nature}), leading to a renewed interest in lattice gauge theories. Simultaneously, it has been realized that constrained systems may also lead to the emergence of constants of motion \cite{ScarsHSF_Moudgalya,DipoleHSF_Khemani,DipoleHSF_Khemani,Moudgalya_2022} which can significantly affect thermalization. One way in which this can occur is through the presence of extensively many conservation laws. This can be in the form of integrability, where the set of conserved quantities completely determines the energy eigenstate, or local Hilbert space fragmentation \cite{LocalHSF_main,LocalHSF_Luitz}, where the determined sectors can in principle be exponentially large. In such cases, thermal equilibrium is no longer described by the Gibbs ensemble, but rather a generalized Gibbs ensemble \cite{GGLoriginal_Rigol,GGLsummary_Rigol}.
    
    In this work, we analyze a wide class of $U(1)$ lattice gauge theories in two dimensions, with a focus on low energy behavior and the emergence of extensively many local, conserved quantities. We focus on the square lattice quantum dimer model (QDM) introduced by Rokhsar and Kivelson (RK) \cite{Original_RK,NotesQDM_Moessner}, and variants such as the QDM on the honeycomb lattice \cite{hxQDM_Moessner}, the quantum six-vertex model \cite{sqQVM_Penc,sqQVM_Moessner}, and a 3-in-3-out spin-ice on the triangular lattice \cite{trQVM_Bannerjee}. The RK models are described by a single independent dimensionless parameter $v$, with ground states that are exactly known at the RK point $v = 1$. When an RK model features a lattice $U(1)$ gauge structure, the RK point can be viewed as a critical spin liquid state with a gapless mode, which separates gapped phases for $v \neq 1$. Determining the precise nature and orders shown by gapped phases for $v < 1$ has been a challenging task in some cases such as the QDM on the square lattice \cite{ED_QDM_Sachdev,NumericsQDM_Runge1996,MixedPhaseQDM_Moessner,QDM_numerics_bannerjee_14,QDM_numerics_bannerjee16}.
    \begin{figure*}
        \centering
        \includegraphics[width = 0.9\textwidth]{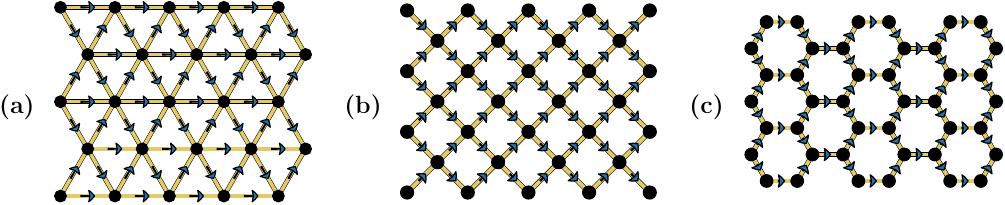}
        \caption{Uniformly polarized configurations for the \textbf{(a)} triangular, \textbf{(b)} square, and \textbf{(c)} hexagonal lattices. The arrows (representing electric field) point from neighboring sites $I$ to $J$ where $I < J$ as defined in the text (See \cref{subsec:general_framework}).}
        \label{fig:intro_lattice}
    \end{figure*}

    A major reason for the difficulty is the fact that the system is two-dimensional, where the capabilities of numerical and analytic approaches remain quite limited. In contrast, a more diverse and powerful arsenal of analytical and numerical tools to tackle and understand interacting systems exists in one dimension such as bosonization, Bethe ansatz, and the density matrix renormalization group (DMRG). This motivates us to obtain effective spin chain models whose behaviors are interesting in their own right, but can also shed light on the key physics of two-dimensional RK models.
    
    Our work will substantially enlarge the approach of Ref. \cite{JonahSebaInti}, which exploited the idea that certain sectors of the full RK model consist of a single extended string-like object, whose dynamics can be mapped to that of a quantum spin chain. Thus, the work of \cite{JonahSebaInti} can be viewed as an investigation of the quantum dynamics of a single non-interacting string (a.k.a electric field line) in a 2D lattice gauge theory. Moreover, the global ground state of the 2D RK model often resides in a sector containing a finite density of closely packed and mutually interacting strings (i.e. many closed electric field lines). To investigate the effects of multiple strings interacting with each other, the present work will devise a series of geometries that allow for string interactions, for example by allowing a string to collide with itself or by considering a pair of closely packed interacting strings. Another of the advancements of our work relative to Ref. \cite{JonahSebaInti}, is that by adding a global discrete height for closed strings, we have found a way to map these onto strictly periodic chains with twisted boundary conditions. Interestingly, we will show that the boundary phase twist maps onto the global transverse momentum of the closed string. This will allow us to show that the standard Drude weight of the 1D chain is exactly equivalent to the inverse of the string mass for its global transverse motions.
    
    Using our understanding of this mapping, we construct about a dozen exact mappings from quasi-one-dimensional RK models to quantum spin-chains. In the models studied, we find that only three distinct one-dimensional spin models arise : (i) the spin-1/2 XXZ chain, (ii) a spin-1 model with global $U(1)$ symmetry, and (iii) a kinetically constrained fermion chain, which we dub the tile chain, due to it admitting an elegant representation in terms of tiles. By performing DMRG simulations of models (ii) and (iii) for moderate system sizes we find, in all but one model, that there are two distinct phases for $v < 1$: a gapped, solid phase and a gapless XY phase. The XY phase can be interpreted as a quasi-long-range ordered one-dimensional descendant of the resonant plaquette phase, and we will show that the spin-1 chain and tile chain show transitions consistent with those reported for the full 2D problems. 
    
    Remarkably we have found that the quantum dimer and six-vertex models that map onto the spin-1 chain display a continuum manifold of exact zero-energy ground states at its RK point analogous to a Bloch sphere, but without an underlying SU(2) symmetry. As a consequence the stiffness of the mode quadratically dispersing gapless mode changes continuously as the "magnetization" of the spin-1 chain is changed.

    Moreover, we find that a particular compactification of the quantum six-vertex model shows several interesting new features. Firstly, the system fragments into multiple disconnected subspaces, with the largest fragment mapping precisely to the tile chain containing the ground state sector, and occupying an exponentially small fraction of the Hilbert space. Secondly, the system shows three distinct phases (unlike all other models we study): A gapped antiferromagnetic (AFM) phase, a gapped resonant plaquette (RP) phase, and a gapless XY phase. Finally, we argue for the existence of a line of Landau-incompatible quantum critical points in the general phase diagram of the tile chain. Our numerical investigation support that the AFM and RP phases of the quasi-1D quantum six-vertex model are indeed separated by a deconfined quantum critical point \cite{OriginalDQCP_Senthil,DQCP_review}.
    
    The paper is organized as follows: Section \ref{sec:gaugedesc} introduces $U(1)$ lattice gauge theories, and explicitly describes the string-descriptions one can obtain from these. Section \ref{sec:single_string} discusses the single string problem with periodic boundary conditions, in the absence of any string-string interactions. Sections \ref{sec:XXZ} and \ref{sec:Spin1} describe mappings onto the spin - 1/2 and spin - 1 chains, with a common phase diagram provided at the end of the respective sections. Section \ref{sec:tile_chain} describes the tile chain, and mappings into it, with a special focus on Landau-forbidden criticality and local Hilbert space fragmentation in the quantum six-vertex problem. We conclude in section \ref{sec:discussion} with a discussion of results and future directions of research.
    
    \section{U(1) Gauge structure of 2D RK hamiltonians}
    \label{sec:gaugedesc}
    
    \subsection{General framework:}
    \label{subsec:general_framework}
    
    \begin{figure*}
        \centering
        \includegraphics[width = 0.9\textwidth]{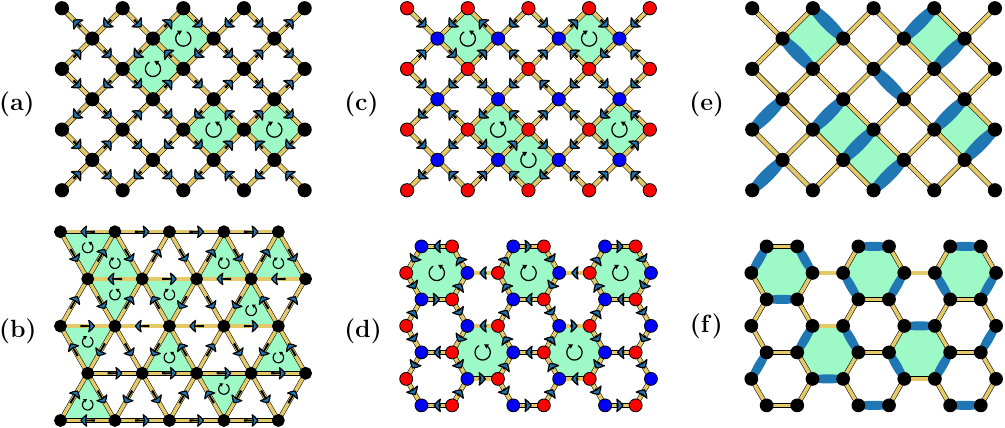}
        \caption{\textbf{(a - d)} Examples of allowed configurations in different $U(1)$ lattice gauge theories, with the plaquettes oriented clockwise ($\circlearrowright$) and anticlockwise ($\circlearrowleft$) marked in green. \textbf{(a)} \sqQVM{}, \textbf{(b)} \trQVM{}, \textbf{(c)} \sqQDM{}, \textbf{(d)} \hxQDM{}. \textbf{(e), (f)} are dimer configurations corresponding to gauge configurations depicted in (c), (d), respectively (see \cref{subsec:QDM}). }
        \label{fig:RepresentativeConfigurations}
    \end{figure*}
    
    We consider $U(1)$ lattice gauge theories defined on the triangular, square, and hexagonal lattices. The degrees of freedom of these models consist of spin - $\frac{1}{2}$ objects present on each link, whose projection along an axis will serve as the lattice versions of the electric field. This notion is made precise by first defining an ordering of the vertices, and subsequently defining a spin - $\frac{1}{2}$ operator $\overrightarrow{\sigma}_{IJ}$ for each link $IJ$ connecting nearest neighbor sites $I < J$. In this work, it shall be convenient to work wfith an ordering of the sites in which $I < J$ if $I$ is ``to the left of'' $J$. If both sites lie on a vertical line, then we say $I < J$ if $I$ is ``below'' $J$ (see Fig.\ref{fig:intro_lattice}). Subsequently, the electric field operator on link $IJ$ with $I < J$ is defined through $E_{IJ} = \sigma^z_{IJ} = - E_{JI}$, which allows us to interpret $E_{I J}$ as the projection of the local electric field vector along the the vector pointing from $I$ to $J$. The configuration with all $\sigma^z_{IJ} = +1$ corresponds to the electric field lines along each link pointing towards the right (except on vertical links, where it points upwards), as illustrated in \cref{fig:intro_lattice}. This configuration will here on be referred to as being uniformly polarized.

    Associated with each vertex $I$ there is a gauge charge $Q_I$ which is defined in analogy with Gauss's law in the continuum, and can be written as the sum of the ``outgoing'' electric field variables as
    \begin{equation}
        Q_I = \sum_{J} E_{IJ}.
    \end{equation}
    The gauge charges can take on a restricted set of values, depending on the lattice. $Q_I \in \{ 0, \pm 2, \pm 4 \}$ for square lattices, $ Q_I \in \{ \pm 1, \pm 3 \} $ for hexagonal lattices, and $ Q_I \pm \{ 0, \pm 2, \pm 4, \pm 6 \}$ for triangular lattices. These charges generate $U(1)$ gauge transformations, denoted $\mathcal{U}$, as
    \begin{equation}
        \mathcal{U} = \exp \left( \ci \sum_{I} \theta_I Q_I \right),
    \end{equation}
    where $\theta_I \in [0, 2 \pi)$ parametrize the transformation.

    A charge hopping (or alternatively, dipole) operator is defined
    \begin{equation}
        p_{IJ} = \begin{cases}
            \sigma^{+}_{IJ} & I < J, \\
            \sigma^{-}_{JI} & I > J,
        \end{cases}
    \end{equation}
    which allows us to define gauge invariant loop operators for a closed path on the lattice $I_1 \rightarrow I_2 \rightarrow \dots \rightarrow I_n \rightarrow I_{n+1} = I_{1}$ as
    \begin{equation}
        L = \prod_{i = 1}^{n} p_{I_i I_{i+1}}.
    \end{equation}
    Loop operators associated with the plaquette $P$ are denoted $L_P$ and $L_P^\dagger$, corresponding to charge hopping operators around anticlockwise and clockwise closed loops respectively, which have been indicated in \cref{fig:RepresentativeConfigurations} (a - d).
    \begin{align}
        L_P &= \ket{\circlearrowleft} \bra{\circlearrowright}_P, \\
        L_P^\dagger &= \ket{\circlearrowright} \bra{\circlearrowleft}_P .
    \end{align}

The Rokhsar-Kivelson (RK) model is a hamiltonian which satisfies the $U(1)$ gauge constraints, and is a sum of terms acting on individual plaquettes:
\begin{equation}
\begin{split}
\label{eqn:RK_general}
    \ham_{\mathrm{RK}} = \sum_P -t ~\kin_P + V ~ \pot_P, \\
\end{split}
\end{equation}
where
\begin{equation}
\begin{split}
    \kin_P &= L_P + L_P^\dagger \\
    &=  \ket{\circlearrowleft} \bra{\circlearrowright}_P + \ket{\circlearrowright} \bra{\circlearrowleft}_P,
\end{split}
\end{equation}
and
\begin{equation}
\begin{split}
    \pot_P &= L_P^\dagger L_P + L_P^\dagger L_P \\ 
    &= \ket{\circlearrowright} \bra{\circlearrowright}_P + \ket{\circlearrowleft} \bra{\circlearrowleft}_P.
\end{split}
\end{equation}
Here $\kin_P$ is the plaquette resonance term, and plays the role of a kinetic term, while $\pot_P$ is diagonal in the $\sigma^z$ basis, and plays the role of a potential term. The ratio $v := V/t$ is the only dimensionless parameter of the hamiltonian and controls the competition between kinetic and potential terms.

RK models are well-defined for several choices of gauge-charges $\{Q_I\}$, which determine different subspaces of the Hilbert space known as sectors of the gauge theory. We will focus on four specific charge configurations: the six-vertex model, and its generalization on the triangular lattice, as well as quantum dimer models on the square and hexagonal lattices.

\subsection{Quantum k-vertex models :}
RK models with gauge charges $Q_I = 0$ on all vertices $I$ of the lattice can be realized on square and triangular lattices, owing to the even coordination number of each site $I$. The model on the square lattice is also known as the quantum six-vertex model \cite{JonahSebaInti} and shall be denoted \sqQVM{}, where it implements the 2-in-2-out rule. This model has recently seen activity in the context of artificial quantum spin ice \cite{qubit_spin_ice,skjaervo_advances_2020}. Analgously, on the triangular lattice we have the 3-in-3-out rule at each vertex \cite{trQVM_Bannerjee}, and shall be denoted \trQVM{}.
\subsection{Quantum dimer models:}
\label{subsec:QDM}
Bipartite lattices allow a configuration of gauge charges which alternate in sign. We will be interested in the case $Q_I = \pm 2$ for the square lattice, and $Q_I = \pm 1$ for the hexagonal lattice. These models can be mapped exactly to quantum dimer models \cite{QDMisGauge_Fradkin,NotesQDM_Moessner,QSFT_Fradkin} on the square and hexagonal lattices, hence we refer to them as \sqQDM{} and \hxQDM{} respectively. This mapping between the gauge theory and the dimer model is demonstrated in \cref{fig:RepresentativeConfigurations}(c, e) for the square lattice. Associated with each vertex $I$ having charge $Q_I = +2$, there exists a unique link $IJ$  where the field points towards $I$. Similarly, for the site $J$, having charge $-2$, $JI$ is the unique link where the field points outwards. Such a link corresponds to a dimer that covers the vertices $I$ and $J$. Thus, given a field configuration, one can determine the corresponding dimer covering in a one-to-one fashion. The analogous construction for hexagonal lattices has been demonstrated in \cref{fig:RepresentativeConfigurations}(d, f).

In terms of dimer configurations, the hamiltonian of \sqQDM{} is 
\begin{equation}
    \begin{split}
        \ham =& -t ~ \sum_P \ket{\dimersquarevert} \bra{\dimersquarehorz} + \ket{\dimersquarehorz} \bra{\dimersquarevert} \\
        &+ V \sum_P \ket{\dimersquarevert} \bra{\dimersquarevert} + \ket{\dimersquarehorz} \bra{\dimersquarehorz},
    \end{split}
\end{equation}
where the sum has been taken over all plaquettes. Analogously, for \hxQDM{} we have
\begin{equation}
    \begin{split}
        \ham =& -t ~ \sum_P \ket{\dimerhexagonup} \bra{\dimerhexagondown} + \ket{\dimerhexagondown} \bra{\dimerhexagonup} \\
        &+ V \sum_P \ket{\dimerhexagonup} \bra{\dimerhexagonup} + \ket{\dimerhexagondown} \bra{\dimerhexagondown}.
    \end{split}
\end{equation}

\subsection{Topological sectors}
\begin{figure}
    \centering
    \includegraphics[width = 0.8\columnwidth]{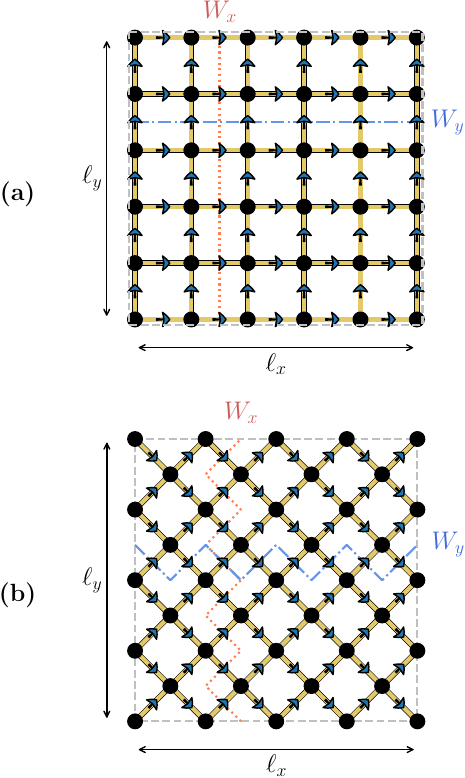}
    \caption{'t Hooft operators $W_x$ and $W_y$ visually depicted for the \sqQVM{} model with periodic boundaries that are \textbf{(a)} aligned with the links \textbf{(b)} at 45$^\circ$ to the links. In both figures, $l_x = l_y = 5$ and the 't Hooft operators for the uniformly polarized configurations are (a) $W_x = 5, W_y = 5$, (b) $W_x = 10, W_y = 0$. }    
    \label{fig:'t Hooft}
\end{figure}

In this work we will largely be interested in RK models compactified onto a torus. Such a compactification gives rise to two additional constants of motion known as 't Hooft operators, denoted $W_x$ and $W_y$, associated with the two non-contractible loops of the torus. For a continuum theory, these operators have the expression
\begin{equation}
    \label{eqn:tHooft_continuum}
    W_{\alpha} = \overrightarrow{n} \cdot \oint_{\alpha^{\perp}} ~ \overrightarrow{dr} \times \overrightarrow{E},
\end{equation}
where the integral is evaluated along a loop along the $y$-direction for $W_x$ and along the $x$-direction for $W_y$, and $\overrightarrow{n}$ is the unit normal. On a lattice, intuitively, these operators count the number of electric field lines going along the $x$ and $y$ directions respectively. This notion can be formalized using \cref{eqn:tHooft_continuum} where instead, a sum over $E$ is performed along a non-contractible path on the dual lattice. An explicit construction can be found in an earlier work \cite{JonahSebaInti}. \cref{fig:'t Hooft} illustrates the operators $W_x$ and $W_y$ for two distinct compactifications of \sqQVM{}.

\subsection{String framework for k-vertex models}
\label{subsec:QVM_strings}
\begin{figure*}
    \centering
    \includegraphics[width = 0.9\textwidth]{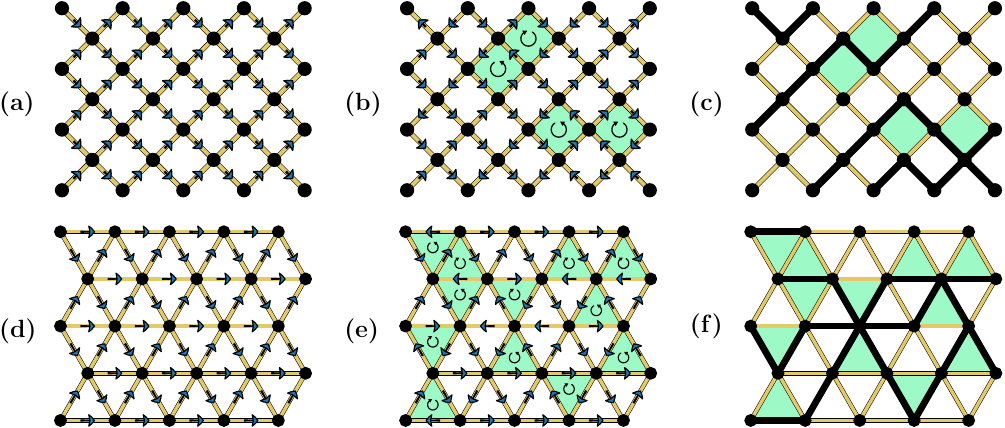}
    \caption{
    \textbf{(a)} Uniformly polarized configuration for the \sqQVM{}, taken to be the configuration with $0$ strings.
    \textbf{(b)} Configuration of \sqQVM{} in terms of fields, which corresponds to 
    \textbf{(c)} in terms of strings. Obtained by marking differences between (a) and (b).
    \textbf{(d)} Uniformly polarized configuration for the \trQVM{}, taken to be the configuration with $0$ strings.
    \textbf{(e)} Configuration of \trQVM{}, in terms of fields, which corresponds to 
    \textbf{(f)} in terms of strings. Obtained by marking differences between (d) and (e).
    }
    \label{fig:QSI_string}
\end{figure*}
An alternative way of looking at configurations and dynamics of the $\overrightarrow{E}$ field is in terms of natural string-like objects. Formally, we define a string to be a continuous path of links where $\sigma^z = -1$, and hence indicate deviations from the uniformly polarized configuration. These strings are guaranteed to be closed by the gauge constraints. Our definition of uniformly polarized configurations further ensures that the strings can be followed from left to right and cannot ``turn around'', as that would again violate the constraints (see \cref{fig:QSI_string}(c), (f)).

In this framework, the 't Hooft operator $W_x$ `counts' the number of strings wrapping around the torus along the $x$ direction. Being a constant of motion, this allows us to divide the Hilbert space into sectors having a different number of strings crossing the (periodic) boundary. Thus, describing the model in terms of its strings makes the $U(1)$ structure manifest, and allows us to discover mappings to simpler models.

The procedure is demonstrated for the \sqQVM{} in \cref{fig:QSI_string}(a-c), and illustrates that the RK hamiltonian written in terms of strings is
\begin{equation}
    \label{eqn:RK_sqQVM}
    \begin{split}
        \ham =& -t ~ \sum_P \ket{\stringsquareup} \bra{\stringsquaredown}_P + \ket{\stringsquaredown} \bra{\stringsquareup}_P \\
        &+ V \sum_P \ket{\stringsquareup} \bra{\stringsquareup}_P + \ket{\stringsquaredown} \bra{\stringsquaredown}_P,
    \end{split}
\end{equation}
where the each term acts on an individual plaquette $P$. The dynamical strings obtained thus have a hardcore repulsion in the sense that no two strings can coincide at a link, which makes the problem intrinsically interacting even for $v = 0$.

An analogous procedure for obtaining a string description of the \trQVM{} is illustrated in \cref{fig:QSI_string}(d-f). For a triangular lattice, we have two distinct kind of plaquettes: those oriented as $\triangle $ and those oriented as $\triangledown$. As a consequence, the hamiltonian has slightly different terms for each kind of plaquette,
\begin{equation}
    \begin{split}
    \ham =& -t ~ \sum_{\triangle} \ket{\stringtriangleupflat} \bra{\stringtriangleupbent} + \ket{\stringtriangleupbent} \bra{\stringtriangleupflat} \\
    &+ V \sum_{\triangle} \ket{\stringtriangleupbent} \bra{\stringtriangleupbent} + \ket{\stringtriangleupflat} \bra{\stringtriangleupflat} \\
    & -t ~ \sum_{\triangledown} \ket{\stringtriangledownflat} \bra{\stringtriangledownbent} + \ket{\stringtriangledownbent} \bra{\stringtriangledownflat} \\
    &+ V \sum_{\triangledown} \ket{\stringtriangledownbent} \bra{\stringtriangledownbent} + \ket{\stringtriangledownflat} \bra{\stringtriangledownflat}.
\end{split}
\end{equation}
More succintly, in terms of the loop operators used to define \cref{eqn:RK_general}, we have 
\begin{equation}
    L_P = \begin{cases}
    ~ \ket{\stringtriangleupbent} \bra{\stringtriangleupflat}  ~ &  ~~  \triangle \textrm{ plaquette } P, \\[0.5em]
    ~ \ket{\stringtriangledownflat} \bra{\stringtriangledownbent} ~ & ~~ \triangledown \textrm{ plaquette } P.
\end{cases}
\end{equation}
The fact that only ``up'' and ``down'' plaquette resonances are admissible for a string follows from the uniformly polarized configuration (\cref{fig:QSI_string}(b)) being used to define the strings. The gauge constraints force the string to have an unbroken trajectory from left to right, which prohibits configurations where the string ``turns back''. 

\subsection{String framework for dimer models}
\label{subsec:QDM_strings}
\begin{figure*}
    \includegraphics[width = 0.95\textwidth]{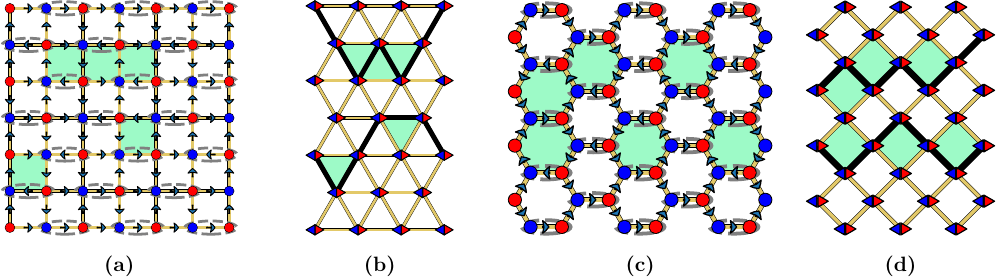}
    \vspace{-0.5em}
    \caption{Mapping of \textbf{(a)} a field configuration in \sqQDM{} to \textbf{(b)} the corresponding string configuration in \trQVM{}, and mapping of \textbf{(c)} a field configuration in \hxQDM{} to \textbf{(d)} the corresponding string configuration in \sqQVM{}. The grouping of vertices used has been shown in (a) and (c). The strings obtained, as in (b) and (d), cannot touch at a composite vertex.}
    \label{fig:StringQDM}
\end{figure*}
For the quantum $k$-vertex models, the gauge charges $Q_I = 0$ allowed an equivalent description in terms of closed strings. Despite the fact that $Q_I \neq 0$ for the dimer models, we can still have a description in terms of string configurations. We define them relative to a staggered dimer configuration (where all the dimers are oriented parallel but no plaquette can resonate) instead of the uniformly polarized configuration. 

We will now describe how every configuration of \sqQDM{} can be mapped onto a configuration of \trQVM{} (but the converse is not true).The similarity with the string framework described in \cref{subsec:QVM_strings} can be made explicit by 
grouping pairs of adjacent vertices, such that the outgoing flux for the group is $0$. For example, in \cref{fig:StringQDM}(a) one can group vertices of \sqQDM{} in a staggered fashion to get a valid configuration of \trQVM{} which can be represented in terms of strings (\cref{fig:StringQDM}(b)) using the procedure described in \cref{subsec:QVM_strings}. 

Although this mapping is injective, not all configurations of \trQVM{} can be obtained from configurations of \sqQDM{}, and thus there is a constraint on the achievable configurations in \trQVM{}. In terms of strings in \trQVM{}, this constraint is ``no two strings can \textit{touch} or coincide at a vertex'' and rules out configurations such as \cref{fig:QSI_string}(f). If this constraint were not there, then strings touching at some vertex in \trQVM{} would also overlap at the link joining their constituents in \sqQDM{}, which cannot occur by our definition of strings. 

In terms of these constrained string configurations of \trQVM{}, the plaquette loop operators used to define the RK hamiltonian (\cref{eqn:RK_general}) are
\begin{equation}
    L_P = \begin{cases}
        ~ \mathcal{P} \ket{\stringtriangleupbentdimer} \bra{\stringtriangleupflatdimer}  \mathcal{P}  ~ &  ~~  \triangle \textrm{ plaquette } P, \\[0.5em]
    ~ \mathcal{P} \ket{\stringtriangledownflatdimer} \bra{\stringtriangledownbentdimer} \mathcal{P}  ~ & ~~ \triangledown \textrm{ plaquette } P,
    \end{cases}
\end{equation}
where $\mathcal{P}$ is the projector onto the subspace of \trQVM{} where no two strings touch. 

Analogously, as shown in \cref{fig:StringQDM}(c, d), there exists a mapping from configurations in \hxQDM{} to constrained string configurations in \sqQVM{}. In terms of these constrained strings of \sqQVM{}, the plaquette loop operator is
\begin{equation}
\begin{split}
    L_P = \mathcal{P} \ket{\stringsquareupdimer} \bra{\stringsquaredowndimer} \mathcal{P},
\end{split}
\end{equation}
where $\mathcal{P}$ again projects onto the subspace where no two strings touch, which excludes configurations such as \cref{fig:QSI_string}(c).

\section{Single string physics}
\label{sec:single_string}
\subsection{Six vertex model - closed boundary conditions}
\label{subsec:closed_strings}

\begin{figure}
    \centering
    \includegraphics[width = 0.7\columnwidth]{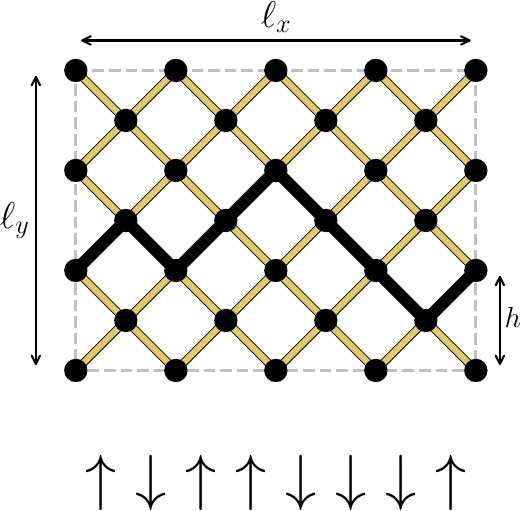}
    \caption{\sqQVM{} compactified onto a torus of dimensions $\ell_x \times \ell_y$ with periodic boundary conditions. The configuration of a single string is mapped onto that of a spin - 1/2 chain, along with a height variable $h$. Here, $\ell_x = 4$, $\ell_y = 3$ and $h = 1$.}
    \label{fig:OneString_PBC}
\end{figure}

In a previous work \cite{JonahSebaInti}, the dynamics of a single string in \sqQVM{} was mapped to an XXZ chain, under the assumption of open boundary conditions. In this section, we will introduce a method to generalize this to the case of  closed strings with periodic boundary conditions. We will see that the momentum along the perpendicular direction to the string periodic direction will play the role of the twist of boundary condition of its effective 1D quantum chain. We revisit those ideas here, in the context of a single string that now wraps around the torus, forming a closed loop. The torus is compactified in the manner shown in \cref{fig:OneString_PBC}, with dimensions $\ell_x$ and $\ell_y$ as indicated. Here we consider the number of strings to be $n_x = 1$, but in general the number of strings is related to the 't Hooft operator as $W_x = W^{\textrm{max}}_x - 2 n_x$, where $W^{\textrm{max}}_x$ is the maximum allowed value of $W_x$.

The configuration of the string can be specified as the sequence of heights of the string $y_i \in \{0, 1/2, 1, \cdots, \ell_y - 1/2\}$ where $i \in \{0, 1, 2, \cdots, \ell_x - 1 \}$ labels the $x$ coordinates from left to right. However, not all $y_i$ are independent and hence it is preferred to describe string configurations using a combination of: (i) The height of the string $h = y_0 \in \{0, 1, \dots \ell_y - 1 \}$ at a fixed reference, taken to be $x = 0$. (ii) The sequence of height changes of the string $\Delta y_i = y_i - y_{i - 1} \in \{ \pm 1/2 \}$, from left to right. Periodic boundary conditions of the torus forces $\sum_i \Delta y_i = 0$, and the Hilbert space of the string is thus $\ell_y \times {2 \ell_x \choose \ell_x }$ dimensional. This description naturally maps onto the configuration of a spin-$1/2$ chain $S^z_i = \Delta y_I$ in the $0$-magnetization sector $\sum_i S^z_i$, along with an extra global ``height'' degree of freedom $h$. 

The RK hamiltonian from \cref{eqn:RK_sqQVM} when acting on a single closed string, maps exactly onto a hamiltonian acting on the corresponding spin-$1/2$ chain as
\begin{equation}
\ham = \ham_{\mathrm{bulk}} + \ham_{\mathrm{edge}},
\end{equation}
where
\begin{equation}
\begin{split}
    \ham_\mathrm{bulk} = & -2t~\sum_{i = 1}^{2\ell_x - 1} (S^{+}_{i+1} S^{-}_{i} + S^{-}_{i+1} S^{+}_{i}) \otimes  \sum_{h = 0}^{\ell_y - 1} \ket{h} \bra{h}   \\
                        & + V~\sum_{i = 1}^{2\ell_x - 1} ( S^{z}_{i+1} - S^{z}_{i} )^2 \otimes  \sum_{h = 0}^{\ell_y - 1} \ket{h} \bra{h},
\end{split}
\end{equation}
and
\begin{equation}
    \label{eqn:edge_flux}
\begin{split}
\ham_\mathrm{edge} = & -2t ~ S^{+}_{1} S^{-}_{2 \ell_x} \otimes  \sum_{h = 0}^{\ell_y - 1} \ket{h + 1} \bra{h} + h.c.   \\
                        & + V ~ ( S^{z}_{2 \ell_x} - S^{z}_{1} )^2 \otimes  \sum_{h = 0}^{\ell_y - 1} \ket{h} \bra{h}.
\end{split}
\end{equation}

\noindent It is evident from the hamiltonian that a translation along the $y$ direction, implemented through a shift in $h$, is a symmetry. The operator which implements this in the spin chain Hilbert space is $T_y(a) = \mathbf{1} \otimes \sum_h \ket{h + a} \bra{h} $, and suggests a change of basis to the $y$-momentum eigenstates $\ket{k_y} = \ell_y^{-1/2}\sum_h e^{ - i k_y h} \ket{h}$. Here $k_y$ is the momentum of the string along the y-direction, and takes values from the set $\{ 0, \pm {2 \pi}/{\ell_y}, \pm 2 \cdot {2 \pi}/{\ell_y}, \cdots \}$, and allows the hamiltonian to be resolved as
\begin{equation}
    \ham = \bigoplus_{k_y} \ham_{k_y} \otimes \ket{k_y} \bra{k_y},
\end{equation}
where
\begin{equation}
\begin{split}
    \ham_{k_y} =& -t \left(e^{\ci k_y} S_0^+ S_{2\ell_x - 1}^- + \sum_{i = 1}^{2\ell_x - 1} S^{+}_{i+1} S^{-}_{i} \right) + h.c.  \\
    &+ V \sum_{i = 1}^{2\ell_x} 2(1 - S_i^z S_{i+1}^z).
    \label{eqn:XXZ_main}
\end{split}
\end{equation}
Thus, the problem decouples into different subspaces with string $y$-momentum $k_y$, with each subspace's hamiltonian corresponding to an XXZ chain with boundary conditions twisted according to $k_y$. This can be interpreted as a problem of hardcore bosons on a ring with flux $\phi = k_y$, suggesting an interesting connection between the dynamics of the string in the $y$-direction and fluctuations in its internal configuration, which we further develop in the next subsection.

\subsection{Mass - Drude weight relation}
When the string is in a gapless phase, the internal fluidity of the string is quantified by the Drude weight $  \mathcal{D} := 2\ell_x \frac{d^2 E} { d \phi^2} \vert_{\phi = 0} $ where $E$ is the ground state energy. Since $k_y$ plays the role of the flux $\phi$, this is proportional to the effective mass of the string along the $y$-direction,
\begin{equation}
    m = \left( \frac{\mathrm{d}^2 E}{\mathrm{d} k_y^2} \right)^{-1} =  \frac{2 \ell_x}{\mathcal{D}}.
\end{equation}
Thus, the more fluid the internal configuration of a string is, the less massive its dispersion in the $y$ direction is. In the thermodynamic limit, although the mass itself tends to infinity one can meaningfully talk about the mass per unit length, which is $2 / \mathcal{D}$.

\subsection{Continuum limit for the fluid string}
To obtain a continuum description for the dynamics of the string, we define a real scalar field $y(x, t)$ denoting the height of the string at coordinate $x$ at time $t$. This field is compact due to the periodic boundary conditions, but restricting ourselves to the $\ell_y \gg \ell_x$ limit, the theory is well described by a non-compact field. We shall assume $v=V/t \in [-1, 1)$ for the present discussion, which in the XXZ model corresponds to the gapless XY phase.

Through Jordan Wigner transformation, the model can be mapped to a 1D spinless fermionic chain. Thus it is possible to describe the continuum physics through a single Luttinger liquid hamiltonian density
\begin{equation}
    \ham(x) = \frac{u}{2} \left( K (\rho(x) - \rho_0)^2 + \frac{1}{K} \pi(x)^2 \right).
\end{equation}
Here $\rho(x)$ is the density of fermions at $x$ and $\rho_0$ is the filling. The coarse grained slope of the string relates to the fermionic density as $\partial y / \partial x = \rho(x) - \rho_0$ and suggests the imaginary-time action for the string-height field $y$ to be
\begin{equation}
    \mathcal{S} = \frac{1}{4 \pi K}\int \mathrm{d}  x ~ \mathrm{d} \tau ~ \frac{1}{u} \left(\frac{\partial y}{\partial \tau} \right)^2 + ~ u \left(\frac{\partial y}{\partial x} \right)^2. 
\end{equation}
For the XXZ model (\cref{eqn:XXZ_main}), the Luttinger parameters $K$ and $u$ are known in terms of the parameters $t$ and $v$ to be
\begin{equation}
\begin{split}
K &= \frac{ \pi }{ \cos^{-1} v }, \\
u &= t ~ \frac{\sqrt{1 - v^2}}{1 - \frac{1}{\pi} \cos^{-1} v}.
\end{split}
\end{equation}

Thus, a single (gapless) string in the thermodynamic limit is described by the Klein-Gordon equation. As a corollary, the fluctuations of the string at finite temperature (in the limit $x \ll \ell_x$) can be described via the correlation functions
\begin{align}
    \langle [y(x,0) - y(0,0)]^2 \rangle &= {K} \log \Big\lvert \frac{\beta u}{\pi \alpha} \sinh \frac{\pi x}{\beta u}    \Big\rvert, \\
    \langle [y(0,t) - y(0,0)]^2 \rangle &= {K} \log \Big\lvert \frac{\beta u}{\pi \alpha} \sinh \frac{\pi t}{\beta}    \Big\rvert.
\end{align}
Here $\alpha$ is the UV regulator and is of the order of the lattice spacing. We see inverse temperature emerging as an effective crossover scale between the limit $t, x/u \ll \beta$, where 
\begin{equation}
\begin{split}
    \langle [y(x,0) - y(0,0)]^2 \rangle &\approx {K} \log \Big\lvert \frac{x}{\alpha}\Big\rvert, \\
    \langle [y(0,t) - y(0,0)]^2 \rangle &\approx {K} \log \Big\lvert \frac{u t}{\alpha}   \Big\rvert,
\end{split}
\end{equation}
and the limit $t, x/u \gg \beta$, where 
\begin{equation}
\begin{split}
    \langle [y(x,0) - y(0,0)]^2 \rangle &\approx K \frac{\pi x}{\beta u}, \\
    \langle [y(0,t) - y(0,0)]^2 \rangle &\approx {K} \frac{\pi t }{\beta }.
\end{split}
\end{equation}
A more complete treatment of the Klein-Gordon equation and its various correlators can be found in \cite{Chaikin_Lubensky,Giamarchi}. Although the effective action of the string has been derived for the \sqQVM{}, we expect this behavior to be fairly generic and independent of the choice of lattice.
\section{Mappings to spin-1/2 XXZ chain}
\label{sec:XXZ}
\begin{figure}
    \centering
    \includegraphics[width = 0.9\columnwidth]{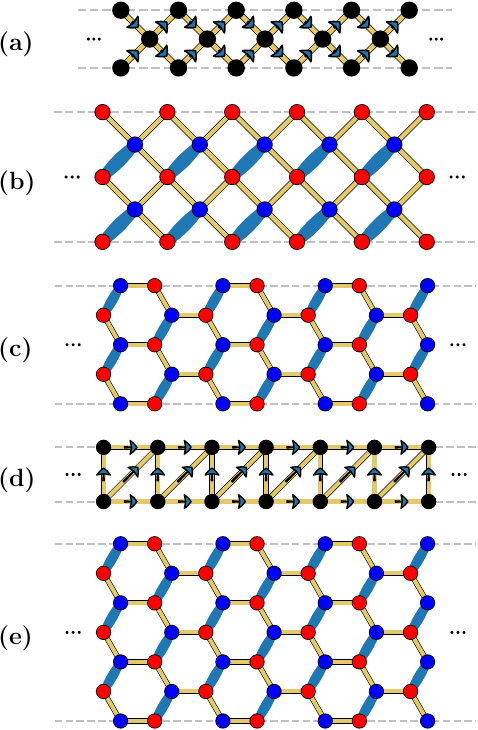}
    \caption{RK hamiltonians defined on these lattices possess sectors that can be mapped to an XXZ chain. Depicted are the zero-string configurations of each model.}
    \label{fig:XXZ_collection}
\end{figure}

In the previous section we described a mapping of a 't Hooft sector containing a single string embedded in a 2D system with arbitrary size. In this section we will map RK Hamiltonians whose y-direction has a small compactification radius but including all its 't Hooft sectors of possibly multiple strings. The lattice geometries for the models that we discuss in this section are shown in \cref{fig:XXZ_collection}, and correspond to the following XXZ hamiltonian:
\begin{equation}
\label{eqn:xxz_zero_flux}
    H_{\mathbf{XXZ}} = -2t \sum_{i = 1}^{ \ell} \left( S^x_i S^x_{i + 1} + S^y_i S^y_{i + 1} + v S^z_i S^z_{i + 1} - v/4 \right).
\end{equation}
Since the models are defined on a torus with dimensions $\ell_x \gg \ell_y \sim 1$, these mappings will naturally follow from their description in terms of strings. In order to accurately capture the ground state degeneracy (GSD) of various phases in the thermodynamic limit, the mapping between $k_y$ and flux in the XXZ chain discussed in \cref{subsec:closed_strings} will be crucial. We use $\mathbf{XXZ}~[\ell, \phi]$ to denote an XXZ chain of length $\ell$ and flux $\phi$.

\subsection{\sqQVM{} - Single non-interacting string}
\label{subsec:XXZ_sqQVM}
\begin{figure}
    \centering
    \includegraphics[width = 0.8\columnwidth]{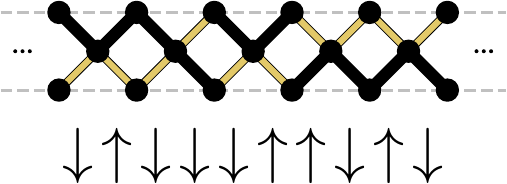}
    \caption{Identification of a single string configuration of \sqQVM{} with that of a spin - half chain.}
    \label{fig:XXZ_sqQVM}
\end{figure}

The associated lattice geometry is depicted in Fig.\ref{fig:XXZ_collection}(a). This is closely related to the discussion in \cref{subsec:closed_strings} with $\ell_y = 1$, and is special in that it allows an analysis of all the three distinct winding number sectors: $n_x = 0, 1,$ and $2$. Of these, the sectors $n_x = 0$ and $n_x = 2$ consist of a single state each and thus are trivial. In the $n_x = 1$ sector, segments of the string oriented along the upper-right direction and those along the lower-right direction are identified with symbols $\uparrow$ and $\downarrow$ respectively (see \cref{fig:XXZ_sqQVM}). Since $T_y$ is identically $1$ for this model, this is an exact mapping to states of a spin-$1/2$ chain. Under this identification, the RK Hamiltonian from \cref{eqn:RK_general} acting on the string maps onto that of an XXZ chain (\cref{eqn:XXZ_main}) of length $\ell = 2 \ell_x$. Thus, the Hilbert space of the model can be partitioned into different sectors as
\begin{equation}
    \mathcal{H} = \bigoplus \begin{cases}
                    \mathbf{0} & n_x = 0 \\
                    \XXZ{2 \ell_x}{0} & n_x = 1 \\
                    \mathbf{0} & n_x = 2.
                \end{cases}
\end{equation}
Sectors with distinct $n_y$ map to sectors of the XXZ model with distinct global $U(1)$ charges.

\subsection{\sqQDM{} - Single self-interacting string}
\label{subsec:sqQDM_single_string}
\begin{figure}
    \centering
    \includegraphics[width = 0.9\columnwidth]{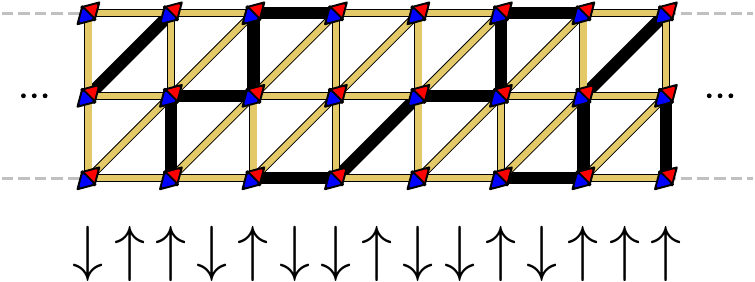}
    \caption{Identification of the single string configuration of \sqQDM{} with that of the XXZ chain. See \cref{subsec:sqQDM_single_string} for further details.}
    \label{fig:XXZ_sqQDM}
\end{figure}

The associated lattice geometry is depicted in Fig.\ref{fig:XXZ_collection}(b). Working about the staggered dimer configuration where the dimers are oriented along the up-right direction, one can group vertices covered by a dimer to obtain an effective \trQVM{} with hardcore string-string repulsion (as discussed in \cref{subsec:QDM_strings} and illustrated in \cref{fig:StringQDM}). An example of such a string configuration is shown in \cref{fig:XXZ_sqQDM}. The constraint of no two strings being allowed to touch restricts the allowed values of the winding number to: $n_x = 0, 1$, and $2$. Of these, sectors $n_x = 0, 2$ possess no dynamics ($\ham$ acts as $0$ on these subspaces), but each sector contributes a $2^{\ell_x}$-fold degeneracy of the zero energy eigenspace.

In the $n_x = 1$ sector, we introduce a spin - $1/2$ degree of freedom associated with each rung (vertical links), as well as a spin - $1/2$ degree of freedom between each rung (horizontal, diagonal links). The mapping from string configurations to spin-chain configurations is according to the rule: follow the trajectory of the string from left to right and if (i) the string covers a vertical link, then assign an $\uparrow$ configuration at that rung; else, assign $\downarrow$; (ii) traversal between rungs via a diagonal link corresponds to an $\uparrow$ configuration there, and $\downarrow$ if a horizontal link is traversed instead. An explicit example of this procedure has been depicted in \cref{fig:XXZ_sqQDM}.

Analogous to the situation in \cref{subsec:closed_strings}, the mapping is not one-to-one and necessitates the inclusion of quantum number $k_y \in \{ 0, \pi \}$ to be complete, which follows from $T_y^{\ell_y} = T_y^2 = 1$. The string hamiltonian then translates exactly to an XXZ spin chain of length $2 \ell_x$, anisotropy $v$, and flux $k_y$.

Thus, the $2^{\ell_x + 1} + 2^{2 \ell_x + 1}$ dimensional Hilbert space of the model can be partitioned into different sectors as
\begin{equation}
    \mathcal{H} = \bigoplus \begin{cases}
                    \mathbf{0}^{\otimes 2 ^{\ell_x}} & n_x = 0 \\[0.2em]
                    \bigoplus_{k_y} \XXZ{2 \ell_x}{k_y} & n_x = 1 \\[0.2em]
                    \mathbf{0}^{\otimes 2 ^{\ell_x}} & n_x = 2.
                \end{cases}
\end{equation}
\subsection{\hxQDM{} - Single self-interacting string}
\label{subsec:hxQDM_single_string}
\begin{figure}
    \centering
    \includegraphics[width = 0.9\columnwidth]{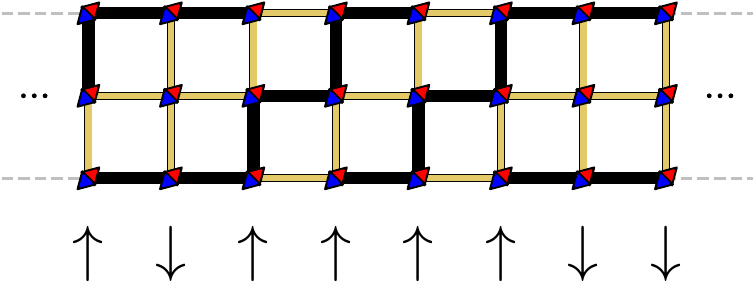}
    \caption{Identification of the single string configuration of \hxQDM{} with that of an XXZ chain. Spin up (down) is assigned when the string does (does not) change its y-position at the given vertex. See \cref{subsec:hxQDM_single_string} for further details.}
    \label{fig:XXZ_hxQDM}
\end{figure}

The \hxQDM{} shown in \cref{fig:XXZ_collection}(c), is compactified onto a narrow torus of length $\ell_x$, and possesses $2 \ell_x$ plaquettes. Forming composite vertices of the up-right oriented vertices yields a string description on the square lattice, which possess two inert sectors corresponding to $n_x = 0$ and $n_x = 2$, with each forming a one-dimensional hilbert space. The $n_x = 1$ sector is mapped to the XXZ chain on $\ell_x$ sites, by identifying segments as shown in \cref{fig:XXZ_hxQDM}, with an additional specification of the $y$-momentum sector $k_y \in \{ 0, \pi \}$, manifesting as an internal flux in the XXZ ring. Thus the $1 + 3 \cdot 2^{\ell_x}$ dimensional Hilbert space of the problem reduces to
\begin{equation}
    \mathcal{H} = \bigoplus \begin{cases}
                \mathbf{0}^{\otimes 2 ^{\ell_x}} & n_x = 0 \\[0.2em]
                \bigoplus_{k_y} \XXZ{\ell_x}{k_y} & n_x = 1 \\[0.2em]
                \mathbf{0}^{} & n_x = 2.
                \end{cases}
\end{equation}
Remarkably, the kinetic and potential terms of the RK hamiltonian map exactly to the hopping and interaction terms of the XXZ chain Hamiltonian from \cref{eqn:xxz_zero_flux}.
\subsection{\trQVM{} - Single self-interacting string}
\label{subsec:trQVM_single_string}
\begin{figure}
    \centering
    \includegraphics[width = 0.8\columnwidth]{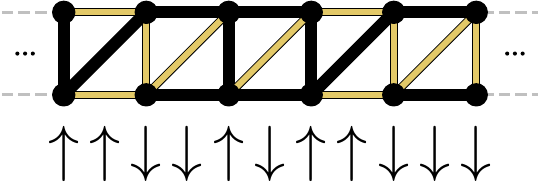}
    \caption{Identification of the single string configuration of \trQVM{} with that of an XXZ chain. See \cref{subsec:trQVM_single_string} for further details.}
    \label{fig:XXZ_trQVM}
\end{figure}

The \trQVM{} shown in \cref{fig:XXZ_collection}(d), is compactified onto a narrow torus of length $\ell_x$, and possesses $2 \ell_x$ triangular plaquettes. For the single string sector $n_x = 1$, one can perform an identification as shown in \cref{fig:XXZ_trQVM} to map the model onto an XZZ chain of $2 \ell_x$ sites. This mapping is exact with no redundancies, as the RK hamiltonian satisfies $T_y = 1$. The remaining sectors $n_x = 0$ and $n_x = 2$ have no dynamics and contribute $2^{\ell_x}$ states each at $0$ energy, implying the Hilbert space breaks up as
\begin{equation}
    \mathcal{H} = \bigoplus \begin{cases}
                    \mathbf{0}^{\otimes 2^{\ell_x}} & n_x = 0, \\[0.2em] 
                    \bigoplus_{k_y} \XXZ{2 \ell_x}{k_y} & n_x = 1, \\[0.2em]
                    \mathbf{0}^{\otimes 2 ^{\ell_x}} & n_x = 2.
                \end{cases}
\end{equation}

\begin{figure}
    \centering
    \includegraphics[width = 0.9\columnwidth]{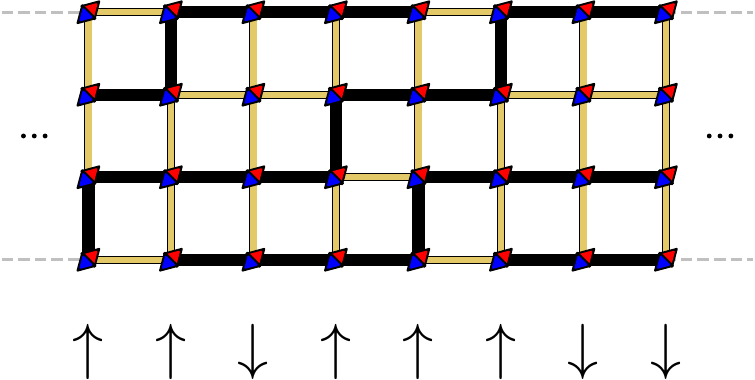}
    \caption{Identification of the two string configuration of \hxQDM{} with that of an XXZ chain. Spin up (down) is assigned when the string does (does not) change its y-position at the given vertex. See \cref{subsec:hxQDM_two_strings} for further details.}
    \label{fig:XXZ_hxQDM_twostring}
\end{figure}

\subsection{\hxQDM{} - Two interacting strings}
\label{subsec:hxQDM_two_strings}
The \hxQDM{} shown in \cref{fig:XXZ_collection}(e) when compactified onto a narrow torus of length $\ell_x$ possesses $3 \ell_x$ plaquettes. While the $n_x = 0$ and $n_x = 3$ sectors are trivial, the sector $n_x = 1$ sector maps onto the spin-1 RK chain, which will be discussed in \cref{subsec:Sp1_hxQDM}. For the two string sector $n_x = 2$, one can perform an identification as shown in \cref{fig:XXZ_hxQDM_twostring} to map the model onto an XZZ chain of $2 \ell_x$ sites \footnote{Notice that the compactification of periodic boundary conditions around the long direction of Fig.\ref{fig:XXZ_hxQDM_twostring} is slightly different compared to that of \cref{fig:XXZ_collection}(e).}. This mapping is three-to-one, with different global string height sectors corresponding to fluxes $k_y = \left\{ 0, \frac{2 \pi}{3}, \frac{4 \pi}{3} \right\}$.

\begin{equation}
    \mathcal{H} = \bigoplus \begin{cases}
        \mathbf{0}^{\otimes 2^{\ell_x}} & n_x = 0 \\[0.2em]
        \bigoplus_{k_y} \SpinOne{\ell_x}{k_y} & n_x = 1 \\[0.2em]
        \bigoplus_{k_y} \XXZ{\ell_x}{k_y} & n_x = 2 \\[0.2em]
        \mathbf{0} & n_x = 3 \\
    \end{cases}
\end{equation}

For energetic reasons (see \cref{appendix:energy_comparison}), the global ground state for all $v < 1$ lies in the sector with $n_x$ = 1, which maps onto the spin-1 RK chain.

\subsection{XXZ features}\label{XXZfeatures}
\begin{figure}
    \centering
    \includegraphics[width = 0.25\columnwidth]{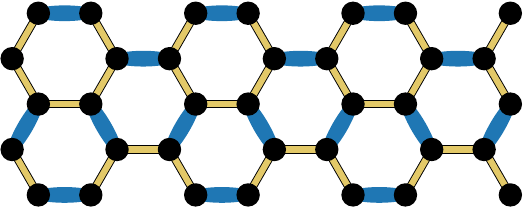} \hspace{0.45\columnwidth}
    \includegraphics[width = 0.25\columnwidth]{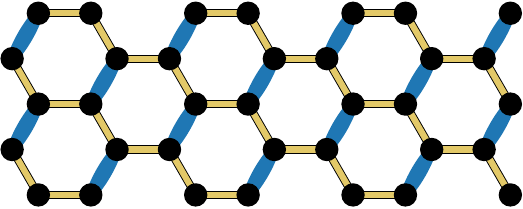} \\ ~ \\

    \includegraphics[width = 0.25\columnwidth]{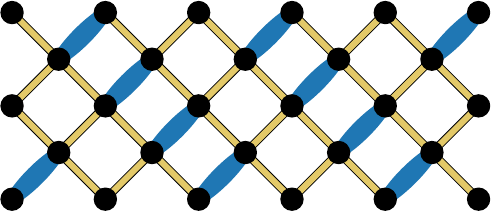} \hspace{0.45\columnwidth}
    \includegraphics[width = 0.25\columnwidth]{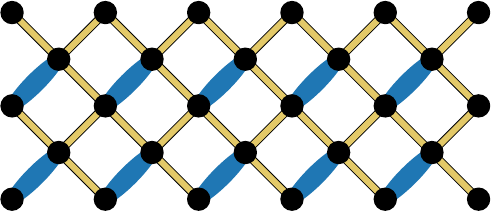} \\ ~ \\

    \includegraphics[width = 0.8 \columnwidth]{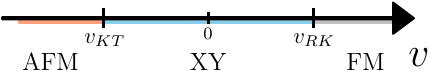}
    \caption{Combined phase diagram for various models described by the XXZ chain. Representative states have been chosen for each solid phase in the lattice gauge theory.}
    \label{fig:pd_XXZ}
\end{figure}

The hamiltonian of the XXZ chain can be expressed in the RK form as
\begin{equation}
    \begin{split}
    H &= -t\sum_I S^+_I S^-_{I + 1} + S^-_I S^+_{I + 1} \\
    & + V \sum_I \left(  S^+_I S^-_{I + 1} \right) \left(  S^-_I S^+_{I + 1} \right) + \left(  S^-_I S^+_{I + 1} \right) \left(  S^+_I S^-_{I + 1} \right), 
\end{split}
\end{equation}
which allows to understand the transition at $v = 1$ as a special form of the RK point sitting at the terminal $SU(2)$ invariant point of the the ferromagnetic phase realized when $v \geq 1$. Furthermore, an exact solution of the spin-$1/2$ XXZ chain is known through Bethe ansatz \cite{XXZ_YangYang1,XXZ_YangYang2}. The chain shows three distinct phases as a function of $v$, with critical points at $v_{KT} = -1$ and $v_{RK} = +1$. The key features of the phase diagram (as shown in \cref{fig:pd_XXZ}) are
\begin{itemize}
    \item $v < -1$: Ising antiferromagnet (gapped, $\GSD = 2$), characterized by the Néel order parameter. 
    \item $ v = -1$: Heisenberg antiferromagnet (gapless, $\omega \sim k$), and also the location of the Kosterlitz-Thouless \cite{Original_KT} transition into the XY phase.
    \item $ -1 \leq v < 1$: XY phase (gapless, $\omega \sim k$), described by a Luttinger liquid in the fermionic representation of the spin chain.
    \item $ v = 1$: Heisenberg ferromagnet (gapless, $\omega \sim k^2$). $\bigotimes \ket{\overrightarrow{m}}$ are the exact ground states, where $\ket{\overrightarrow{m}}$ is an $SU(2)$ coherent state, $ ( \overrightarrow{m} \cdot \overrightarrow{S} ) \ket{\overrightarrow{m}} = \ket{\overrightarrow{m}}$. Displays a first order transition. 
    \item $1 \leq v$: Ising ferromagnet (gapped, GSD = 2). $\bigotimes \ket{\uparrow}$ and $\bigotimes \ket{\downarrow}$ are the exact ground states of the system.
\end{itemize}

The quadratic low energy dispersion $\omega \sim k^2$ of the RK point will be displayed by the spin-1 RK chain. It has been shown under the single mode approximation \cite{QSFT_Fradkin} that this feature persists in the $2D$ problem as well. We therefore see that the spin-fluid (XY) phase obtained in intermediate regime, can be naturally interpreted as a quasi-long range ordered descendant of the resonant plaquette phase believed to exist in the corresponding 2D RK models. It is interesting to note that this duality establishes each of the models in question to be integrable through this mapping. 
\section{Mappings to spin-1 RK chain}
\label{sec:Spin1}
\begin{figure}
    \centering
    \includegraphics[width = 0.9\columnwidth]{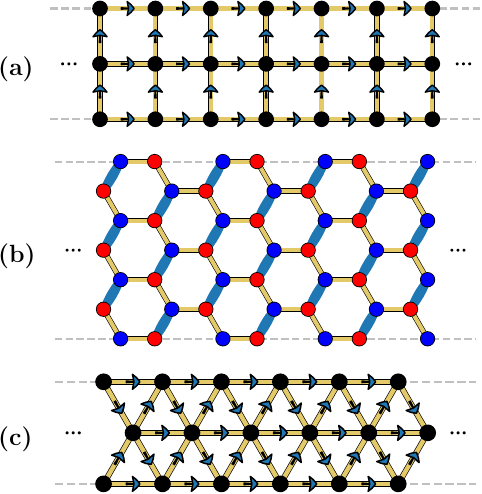}
    \caption{Class of models which map to the spin-one chain. \textbf{Top to Bottom:} figures labeled (a) to (c).}
    \label{fig:Sp1_collection}
\end{figure}

In this section, we map certain sectors of the three models shown in \cref{fig:Sp1_collection} to a one-dimensional spin-1 hamiltonian, which we dub the spin-1 RK chain, given by
\begin{equation}
\label{eqn:spinone_rkchain_hamiltonian}
\begin{split}
    \ham_{\textrm{Sp-1}} =& - t \sum_{I} S_I^x S_{I + 1}^x + S_I^y S_{I + 1}^y \\
    &+ V \sum_{I} f(S_I^z, S_{I+1}^z).
\end{split}
\end{equation}
Where $S^{x,y,z}_I$ are standard spin-1 operators, and the potential term $f$ can be written in the $S^z$ basis as
\begin{equation}
    f(S_I^z, S_{I+1}^z) = \ket{00}\bra{00} - \left( \ket{\uparrow \uparrow}\bra{\uparrow \uparrow} + \ket{\downarrow \downarrow }\bra{\downarrow \downarrow} \right),
\end{equation}
where $\ket{\uparrow}, \ket{0}$ and $ \ket{\downarrow}$ are the eigenstates of $S^z$ with eigenvalues $+1, 0$ and $-1$ respectively. The GSD of various RK models will again be captured by copies of spin chains with different (twisted) boundary conditions. We define the flux $\phi$ by adjusting the kinetic term on the last bond in PBC,
\begin{equation}
    \ham^{\textrm{kin}, 1 -L}_{\textrm{Sp-1}}= -\frac{t}{2} \left( e^{\ci \phi} S_{L}^{+} S_{1}^{-} + e^{-\ci \phi} S_{L}^{-} S_{1}^{+} \right).
\end{equation}

\subsection{\sqQVM{} - Single self-interacting string}
\begin{figure}
    \centering
    \includegraphics[width = 0.8\columnwidth]{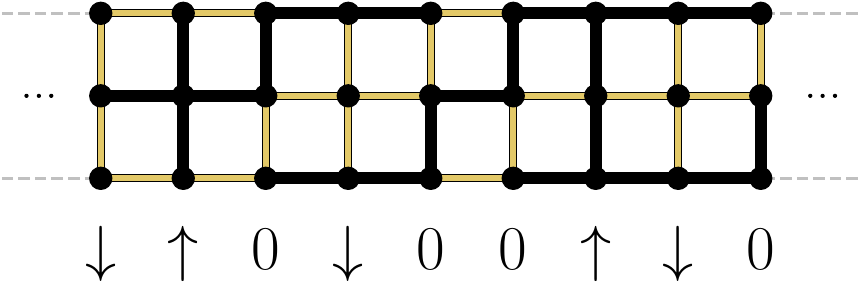}
    \caption{Identification of the single string configuration of \sqQVM{} with a spin-1 RK chain.}
    \label{fig:Sp1_sqQVM}
\end{figure}

The \sqQVM{} on a narrow torus of length $\ell_x$ as shown in \cref{fig:Sp1_collection}(a) possesses $2 \ell_x$ plaquettes and can be divided into sectors $n_x = 0, 1, 2$ based on number of strings. We point out this model has been studied before \cite{Spin1RK_Banerjee}, and our spin-chain mapping provides insight into the results obtained there. The sectors $n_x = 0$ and $n_x = 2$ are trivial ($H$ acts as $\mathbf{0}$), and contribute a degeneracy of $2^{\ell_x}$ each to the zero energy manifold. The Hilbert space as well as the hamiltonian of the $n_x = 1$ sector can be identified with the spin-one model by mapping vertical rung $I$ containing $n$ segments of the string with $\ket{S_I = n - 1}$, as demonstrated in \cref{fig:Sp1_sqQVM}. This identification requires in addition the specification of the $y$-momentum sector $k_y \in \{0, \pi \}$, leading to the decomposition of the Hilbert space

\begin{equation}
    \mathcal{H} = \bigoplus
    \begin{cases}
        \mathbf{0}^{2^\ell_x}                   & n_x = 0, \\[0.2em]
        \bigoplus_{k_y} \SpinOne{\ell_x}{k_y}  & n_x = 1, \\[0.2em]
        \mathbf{0}^{2^\ell_x}                   & n_x = 2.
    \end{cases}
\end{equation}

\subsection{\hxQDM{} - Single self-interacting string}
\label{subsec:Sp1_hxQDM}
\begin{figure}
    \centering
    \includegraphics[width = 0.8\columnwidth]{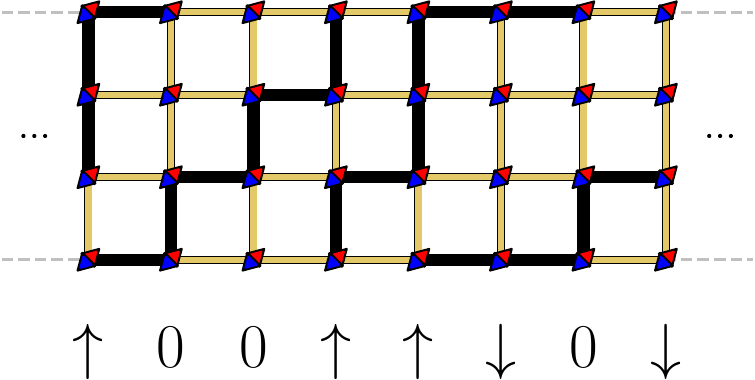}
    \caption{Identification of the single string configuration of \hxQDM{} with the spin - 1 RK chain.}
    \label{fig:Sp1_hxQDM}
\end{figure}

The \hxQDM{} shown in \cref{fig:Sp1_collection} (b), is compactified onto a narrow torus of length $\ell_x$ , and possesses $3 \ell_x$ plaquettes. Performing dimer contraction along the links oriented along the top-right direction, we obtain an equivalent string model on the square lattice. The sector $n_x = 2$ has been discussed in \cref{subsec:XXZ_hxQDM}, and maps to the XXZ chain. Sector $n_x = 0$ has a dimension of $2^{\ell_x}$, and possesses no dynamics, while the sector $n_x = 3$ contains a unique state with no dynamics. The sector $n_x = 1$ can be identified with the spin-one model using identifications shown in \cref{fig:Sp1_hxQDM}, where the hardcore interaction constraint of the string restricts configurations where the string touches itself. This identification is incomplete unless we include the $y$-momentum sectors $k_y = 0, 2 \pi/3, 4 \pi/3$. Thus, the Hilbert space gets broken up as
\begin{equation}\label{Hsp1}
    \mathcal{H} = \bigoplus \begin{cases}
        \mathbf{0}^{\otimes 2^{\ell_x}} & n_x = 0 \\[0.2em]
        \bigoplus_{k_y} \SpinOne{\ell_x}{k_y} & n_x = 1 \\[0.2em]
        \bigoplus_{k_y} \XXZ{\ell_x}{k_y} & n_x = 2 \\[0.2em]
        \mathbf{0} & n_x = 3 \\
    \end{cases}
\end{equation}
Of these sectors, the ground state typically corresponds to the $\mathbf{Spin - 1} (k_y = 0 )$ chain, as we shall see in \cref{subsec:spinone_numericsanddiscussion}.

\subsection{\trQVM{} - Two interacting strings}
\label{subsec:QVM_triangle}
\begin{figure}
    \centering
    \includegraphics[width = 0.8\columnwidth]{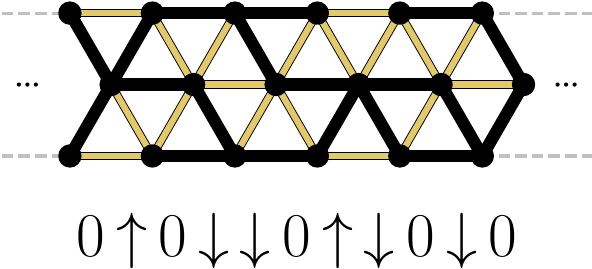}
    \caption{Identification of the two-string sector of \trQVM{} with the spin - 1 RK chain.}
    \label{fig:Sp1_trQVM}
\end{figure}

\begin{figure}
    \includegraphics[width = 0.7\columnwidth]{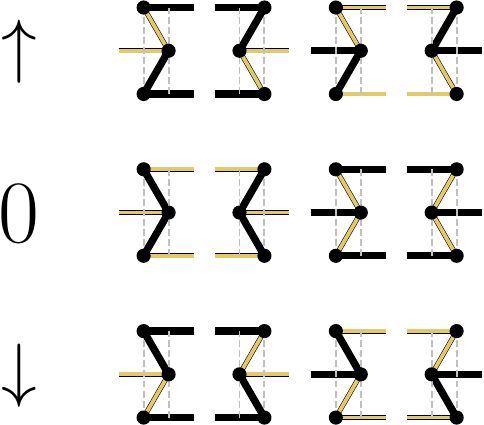}
    \caption{Dictionary for mapping two-string configurations of \trQVM{} (\cref{fig:Sp1_collection}(c)) to configurations of a spin-$1$ chain in the $S^z = +1, 0, -1$ basis.}
    \label{fig:SpinOneDict}
\end{figure}

The \trQVM{} on a narrow torus of length $\ell_x$ as shown in \cref{fig:Sp1_trQVM} with $4 \ell_x$ plaquettes, and can be divided into sectors $n_x = 0, 1, 2, 3, 4$ based on the number of strings (relative to the uniform polarization). The sectors $n_x = 0$ and $n_x = 4$ are trivial, with a unique state in each with no dynamics. The sectors $n_x = 1$ and $n_x = 3$ each map to the tile chain (see \cref{subsec:QVM_triangle}). The sector $n_x = 2$ maps to the spin-one RK chain based on the identifications described in \cref{fig:SpinOneDict} and explicitly demonstrated in \cref{fig:Sp1_trQVM}. Specification of $y$-momentum sectors makes the mapping one-one, and captures the expected ground state degeneracies faithfully. The RK-hamiltonian maps exactly to the Spin-1 hamiltonian under this mapping, and allows us to decompose the Hilbert space as
\begin{equation}
    \mathcal{H} = \bigoplus
    \begin{cases}
        \mathbf{0}                                  & n_x = 0 \\[0.2em]
        \bigoplus_{k_y} \TileChain{2 \ell_x}{k_y}  & n_x = 1 \\[0.2em]
        \bigoplus_{k_y} \SpinOne{2 \ell_x}{k_y}    & n_x = 2 \\[0.2em]
        \bigoplus_{k_y} \TileChain{2 \ell_x}{k_y}  & n_x = 3 \\[0.2em]
        \mathbf{0}                                  & n_x = 4
    \end{cases}
\end{equation}
For all $v < -1$, the energetically lowest state happens to lie in the spin-1 RK chain sector, which we confirm numerically in \cref{appendix:energy_comparison}.

\subsection{Properties of Spin-1 RK chain}
\label{subsec:spinone_numericsanddiscussion}

The hamiltonian from \cref{eqn:spinone_rkchain_hamiltonian} has a global $U(1)$ symmetry generated by the charge $\sum_I S_I^z$, which is a remnant of the $U(1)$ gauge structure of the full $2D$ model. Time reversal is a symmetry and is implemented $e^{\ci \pi \sum_I S_I^Y} \mathcal{K}$, where $\mathcal{K}$ is a conjugation operator in the $S^Z$ basis. In addition, $e^{\ci \pi \sum_I S_I^X}$ is a global $\mathbb{Z}_2$ symmetry which maps $S^z \rightarrow -S^z$.

\subsubsection{Phase diagram}

\begin{figure}
    \centering
    \includegraphics[width = 0.25\columnwidth]{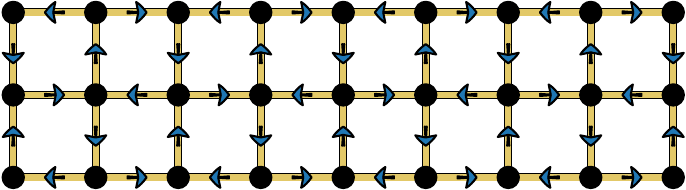} \hspace{0.3\columnwidth}
    \includegraphics[width = 0.25\columnwidth]{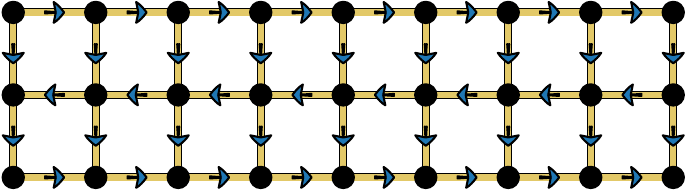} \\ ~ \\

    \includegraphics[width = 0.25\columnwidth]{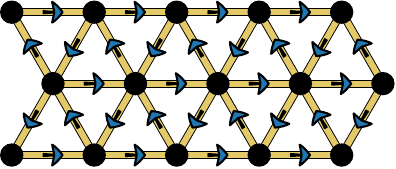} \hspace{0.3\columnwidth}
    \includegraphics[width = 0.25\columnwidth]{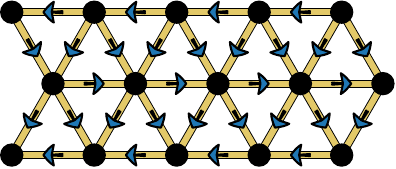} \\ ~ \\

    \includegraphics[width = 0.25\columnwidth]{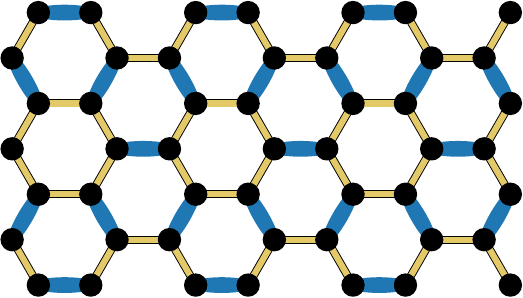} \hspace{0.3\columnwidth}
    \includegraphics[width = 0.25\columnwidth]{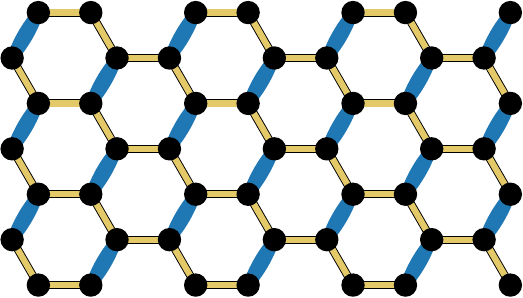} \\ ~ \\

    \includegraphics[width = 0.9 \columnwidth]{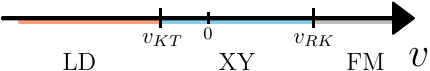}
    \caption{Combined phase diagram for various models described by the spin-1 RK chain. Representative states have been chosen for each solid phase in the lattice gauge theory.}
    \label{fig:pd_SpinOne}
\end{figure}

The phase diagram we have determined via DMRG analysis of the model with open boundary conditions is summarized in Fig.\ref{fig:pd_SpinOne}. It features the following phases and critical points that we summarize below and will further explain in the coming sub-sections:

\begin{enumerate}
    \item  $v < v_{KT}$: (Gapped, GSD = 1). Is a quantum paramagnetic phase which evolves adiabatically from the product state $ \otimes_I \ket{0}_I$, and also referred to as the large-D (LD) phase \cite{LargeD_S1_XXZ}.
    \item $v = v_{KT}$: (Gapless, $\omega \sim k$) Location of a Kosterlitz-Thouless (KT) phase transition point into the XY phase. We numerically estimate $v_{KT} \sim -0.4$ (see \cref{fig:Spin1_Numerics}).
    \item $v_{KT}< v < v_{RK}=1$: (Gapless. $\omega \sim k$) XY magnetic phase characterized by central charge $c = 1$.
    \item $v = v_{RK}=1$: (Gapless, $\omega \sim k^2$) RK point realizing continuum of zero energy ground states without global $SU(2)$ symmetry, and located first order phase transition into the ferromagnet (see below and contrast with RK point in XXZ case from Sec.\ref{XXZfeatures}).
    \item $v > v_{RK}=1$: (Gapped, GSD = 2) Ferromagnetic phase, $\bigotimes_I \ket{\uparrow}$ and $\bigotimes_I \ket{\downarrow}$ are the two exact ground states for all points in this region.
\end{enumerate}

\subsubsection{RK point and low energy excitations}
\noindent The hamiltonian from Eq.\eqref{Hsp1} can be re-written as follows:
\begin{equation}
\begin{split}
    H =& -t\sum_I \frac{1}{2} S^+_I S^-_{I + 1} + \frac{1}{2} S^-_I S^+_{I + 1} \\
    & +v \sum_I \left( \frac{1}{2} S^+_I S^-_{I + 1} \right) \left( \frac{1}{2} S^-_I S^+_{I + 1} \right) + \left( \frac{1}{2} S^-_I S^+_{I + 1} \right) \left( \frac{1}{2} S^+_I S^-_{I + 1} \right) 
\end{split}
\end{equation}
which is of the RK form (see \cref{eqn:RK_general}) with $L_I = S^+_I S^-_I$, and hence reduces to a sum of projectors at the point $v = 1$ (see \cref{appendix:spinone_sumofprojectors}). At this point $v = 1$, we have found a continuum of ground states in the infinite system, which are spanned by the product states of the form:
\begin{equation}\label{RK-zstates}
    \ket{GS (z)} = \bigotimes_I \ket{z}_I
\end{equation}
where the single-spin state is specified by a $z \in \mathbb{C} \cup \{\infty \}$ and is given by:
\begin{equation}
     \ket{z} = \frac{\ket{\uparrow} + z\ket{0} + z^2 \ket{\downarrow}}{\sqrt{1 + |z|^2 + |z|^4}}.
 \end{equation}
Notice that the above is not the $SU(2)$ spin-1 coherent state (which would contain instead the coefficient ket $\sqrt{2}z$ in the spin zero ket), but is a different coherent manifold of the same dimensionality. Remarkably, by performing the linearized analysis of collective modes, we have also found that the stiffness of the quadratically dispersing mode changes continuously as a function of magnetization, and their low-energy dispersion is given by (see Appendix \ref{appendix:path_integral}):

\begin{equation}
    \omega(k) = t \frac{(1 + |z|^2)^2}{1 + 4 |z|^2 + |z|^4} k^2
\end{equation}

\noindent The softest dispersion is realized for the zero magnetization space with $|z|=1$. This demonstrates directly that, in sharp contrast to the XXZ chain from Sec.\ref{XXZfeatures}, this manifold of exact translationally invariant, degenerate ground states is not generated by any global microscopic symmetry. 
This lack of symmetry protection of the stiffness, is expected to lead to an order by disorder effect for $T > 0$, where thermal fluctuations about states with $|z| = 1$ lower the free energy of the $0$ - magnetization sector.

\subsubsection{Large D phase and the even-string order parameter}
The gapped phase at $v < v_{KT}$ breaks none of the symmetries and is topologically trivial, and hence can be characterized using the even-string order parameter introduced in \cite{1DSPT_Pollman},
\begin{equation}\label{evenO}
   O_{\textrm{even}}^{i, j} =  \bra{\psi} \prod^{j - 1}_{k = i + 1} e^{\ci \pi S^z_k} \ket{\psi}. 
\end{equation}
This is the trivial-phase counterpart of the odd-string order parameter \cite{StringOrderParameter_Nijs,StringOrderParameter_Cirac} used to characterize the spin-1 Haldane phase \cite{HaldanePhase1_Haldane,HaldanePhase2_Haldane,HaldanePhase_Tasaki}. The even string order parameter \footnote{\href{https://www.itp3.uni-stuttgart.de/downloads/theses/MasterThesis_JohannesMoegerle_2022.pdf}{\color{gray}{Johannes Moegerle, Master's Thesis, University of Stutgart}}} provides a positive test for the large-$D$ phase \cite{LargeD_S1_XXZ}, where $D$ conventionally refers to the coefficient of the single-ion anisotropy term $ D \sum_{I} \left(S_I^z\right)^2$.

\begin{figure}
    \centering
    \includegraphics[width=\columnwidth]{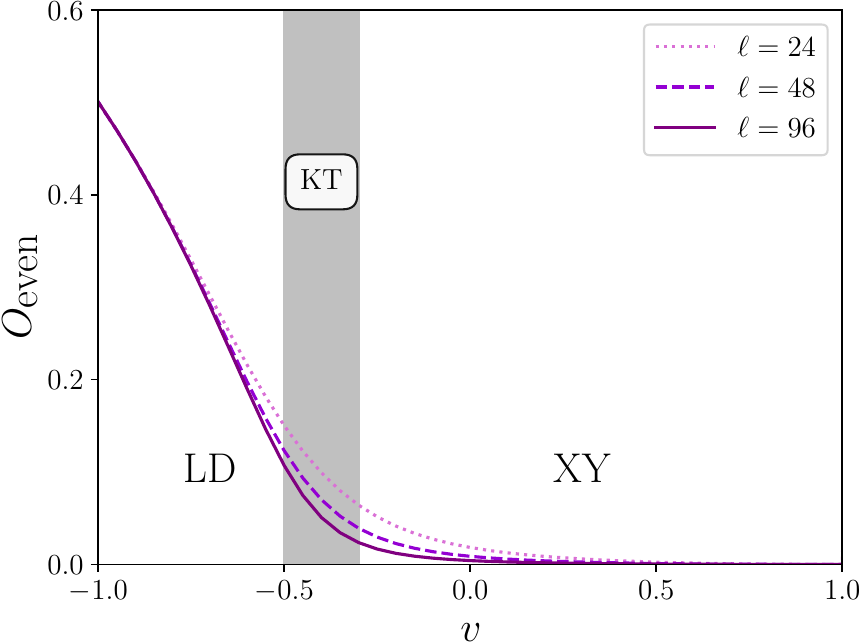} \\ ~ \\
    \includegraphics[width=\columnwidth]{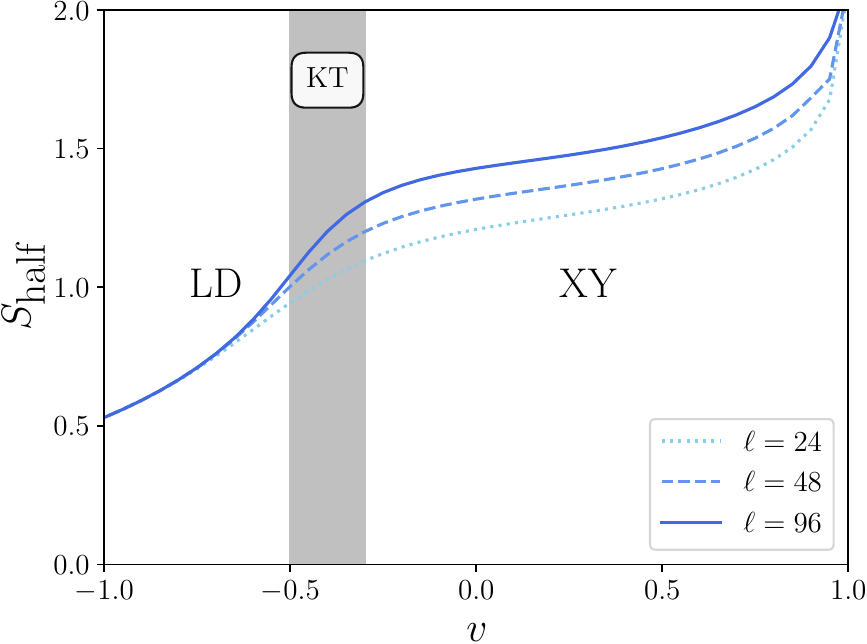}
    \caption{\textbf{Top:} Even-string order parameter (see Eq.\eqref{evenO}) measured from between $i = \ell/4$ to site $j = 3 \ell / 4$ in an open chain of size $\ell$. \textbf{Bottom:} Half-cut entanglement entropy in an open chain.}
    \label{fig:Spin1_Numerics}
\end{figure}

\subsubsection{XY phase}
The XY phase is characterized by its central charge $c = 1$, which we have probed in DMRG via the Calabrese-Cardy \cite{Calabrese_Cardy_Holzhey,Calabrese_Cardy_QFT,Calabrese_Cardy_CFT} formula for the entanglement entropy for a subsystem of the chain, 
\begin{equation}
    S = \frac{c}{6} \log \left( \frac{L}{2 \pi} \sin \frac{\pi l}{L} \right) + \cdots.
\end{equation}
It is favorable to use the entanglement entropy at the half-cut to extrapolate the central charge
\begin{equation}
    S_{\textrm{bipartite}} = \frac{c}{6} \log{L} + \cdots,
\end{equation}
where the additive terms are non-universal and subleading. In the bottom panel of Fig.\ref{fig:Spin1_Numerics} we show that for $-0.5 <\sim v <\sim1$, every doubling of the system size leads to the same constant shift of the entanglement entropy consistent with $c=1$, supporting the claim that we have XY phase between $v_{KT}$ and $v_{RK}$.

\section{Mappings to the tile chain}
\label{sec:tile_chain}
\begin{figure}
    \centering
    \includegraphics[width = 0.8\columnwidth]{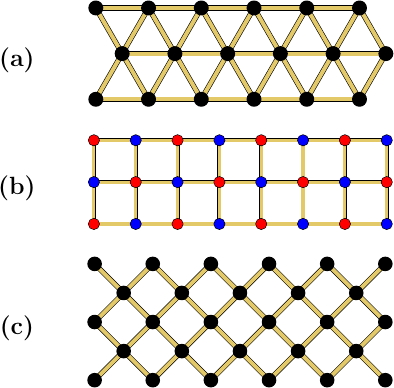}
    \caption{Collection of models which reduce to the tile chain in specific sectors.}
    \label{fig:TT_collection}
\end{figure}

In the models we have considered so far (\cref{fig:XXZ_collection,fig:Sp1_collection}), the relevant sectors can be mapped onto either a spin-$1/2$ or spin-$1$ chain. That is, the resultant Hilbert space is a tensor product of local Hilbert spaces, which correspond to spin-$1/2$ or spin-$1$ local degrees of freedom. Typically, even though the underlying full Hilbert of a generic $U(1)$ lattice gauge theory on a torus is a tensor product, sub-spaces with definite gauge charge configuration and winding number need not be expressible as a tensor product, making such mappings hard to generalize. However, for a certain class of such models (see \cref{fig:TT_collection}), we have found mappings to what we dub a `tile chain', where many body configurations are labeled by tilings of a strip. 
These configurations can alternatively be viewed in terms of spinless fermions with additional near-neighbor hardcore occupation constraints (to be discussed in \cref{hardcore-Rep}).

In the tile chain, a configuration of length $\ell$ is defined as an arrangement of tiles $\tileA,~\tileB,~\tileS$ of lengths $1,1$ and $2$ (respectively) end to end, such that the lengths add up to $\ell$. Some examples of configurations are illustrated in \cref{fig:tile_configs}. The Hilbert space of a tile chain of length $\ell$ is defined as the complex vector space spanned by the orthonormal ket vectors corresponding to each distinct configuration of length $\ell$. In general, the Hilbert space dimension can be shown to grow as $(1 + \sqrt{2})^\ell$ (see \cref{appendix:hilbertspacefragmentation}), demonstrating that there is no one-to-one mapping to a simple tensor product space. In \cref{hardcore-Rep}, we shall discuss an embedding of the model in a larger tensor product space of spin-less fermions, by imposing hardcore constraints on neighboring occupations and allowing an interpretation of the various phases as charge density waves. When $\ell_x = 2$, the Hilbert of the model with open boundary conditions is of dimension $5$ and spanned by the configuration kets
\begin{equation*}
\begin{split}
\ket{\hspace{0.004\columnwidth}\tileA \tileA}, \ket{\hspace{0.004\columnwidth}\tileA \tileB}, \ket{\hspace{0.004\columnwidth}\tileB \tileA}, \ket{\hspace{0.004\columnwidth}\tileB \tileB} ~\textrm{and}~ \ket{\hspace{0.004\columnwidth}\tileS}.
\end{split}
\end{equation*}

The respective hamiltonian of the tile chain is a sum of terms which act only on positions $I$ and $I+1$, where $I \in \{ 1, 2, \ldots \ell_x \}$. The dynamic (kinetic) term is
\begin{equation}
\begin{split}
    \hat{K}_I &= \ket{\hspace{0.004\columnwidth}\tileS}_{I, I+1} \bra{\tileA \tileB}_{I, I+1} + h.c. \\
    &+ \ket{\hspace{0.004\columnwidth}\tileS}_{I, I+1} \bra{\tileB \tileA}_{I, I+1} + h.c.
\end{split}
\end{equation}
while there are two kinds of potential terms,
\begin{equation}
\begin{split}
    \hat{P}^{AB}_I &= \ket{\hspace{0.004\columnwidth}\tileA \tileB}_{I, I+1} \bra{\tileA \tileB}_{I, I+1} \\
    &+ \ket{\hspace{0.004\columnwidth}\tileB \tileA}_{I, I+1} \bra{\tileB \tileA}_{I, I+1}
\end{split}
\end{equation}
and
\begin{equation}
\begin{split}
    \hat{P}^{S}_I = \ket{\hspace{0.004\columnwidth}\tileS}_{I, I+1} \bra{\tileS}_{I, I+1}.
\end{split}
\end{equation}
The hamiltonian for a length $\ell$ tile chain can thus be expressed
\begin{equation}
\label{eqn:TileModel}
    \hat{H} = \sum^{\ell - 1}_{I = 0} -t ~ \hat{K}_I + \VAB ~ \hat{P}^{AB}_I + \VS ~ \hat{P}^{S}_I.
\end{equation}
Denote $v_{AB} = \VAB/t$ and $v_{S} = \VS/t$ as the two dimensionless parameters in the theory. We will now show mappings from RK models with parameter $v = V/t$ to the tile chain with parameters $v_{AB}, v_{S}$, and subsequently discuss the general phase diagram of the tile chain.
\begin{figure}
    \centering
    \begin{align*}
        \tileA \tileB \tileS \tileA \tileS \tileB \tileS \tileS \\
        \tileS \tileS \tileS \tileS \tileS \tileS \\
        \tileS \tileA \tileA \tileS \tileA \tileA \tileS \tileA \tileB  \\
        \tileA \tileB \tileB \tileA \tileA \tileB  \tileB \tileB \tileA \tileA \tileA \tileB
    \end{align*}
    \caption{Four distinct tile configurations of length $\ell_x = 12$.}
    \label{fig:tile_configs}
\end{figure}
\subsection{\trQVM{}}
\label{subsec:tiles_QVMtriangular}
Sectors $n_x = 0, 2 $ and $4$ of the model have been discussed in \cref{subsec:QVM_triangle}. For the sector $n_x = 1$, one can map the up-right segments of the string to $\tileA$, the down-right segments of the string to $\tileB$, and the horizontal segments to $\tileS$ as shown in \cref{fig:TT_QDM}. Under this identification, the RK hamiltonian maps to the tile chain with parameters $\VS = 2 \VAB = 2 V$ and hopping parameter $t$.

The sector $n_x = 3$ is in a one-to-one correspondence with the sector $n_x = 1$, and hence, is also described by the tile chain with parameters $\VS = 2 \VAB = V$. This correspondence is obtained by flipping the spins on all links (achieved by the action of operator $\prod_l \sigma_l^x$) as demonstrated in \cref{fig:TT_QDM}.

\subsection{\sqQDM{}}
\begin{figure}
    \centering
    \includegraphics[width = 0.84\columnwidth]{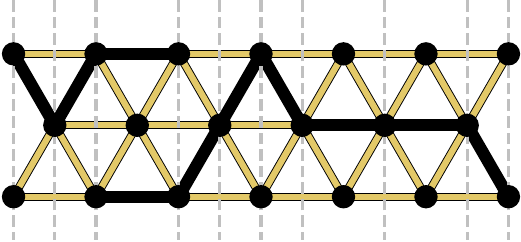}
    \includegraphics[width = 0.8\columnwidth]{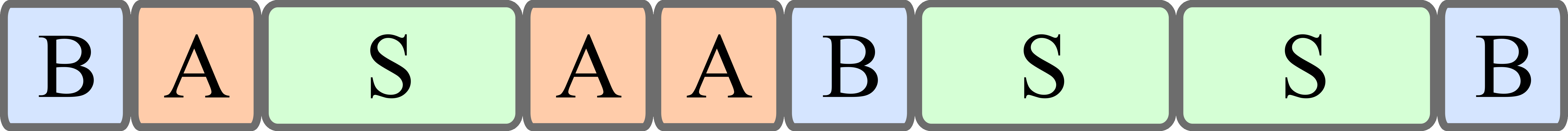}
    \includegraphics[width = 0.84\columnwidth]{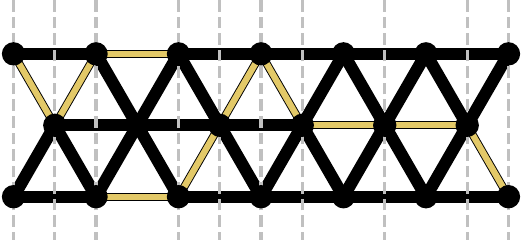}
    \caption{Identification of the one-string and three-string sectors of \trQVM{} with the tile model.}
    \label{fig:TT_QDM}
\end{figure}

The \sqQDM{} (\cref{fig:TT_collection}(b)) has length $\ell_x$, and maps to a string problem on the \trQVM{} with hardcore string-string repulsion. This is similar to the QDM ladder considered in \cite{QDMladder_Moessner,QDMladder_DMRG}, but the presence of periodic boundary conditions in the $y$-direction leads to qualitatively different behaviors. There are three distinct 't Hooft sectors $n_x = 0, 1$ and $2$, Of which, $n_x = 0$ and $n_x = 2$ are trivial and contain a single state each with no dynamics. The sector $n_x = 1$ corresponds exactly to the single string sector of \trQVM{} as shown in \cref{fig:TT_QDM}(c), and is analogously described by the tile chain with $\VS = 2 \VAB = 2V$.

\subsection{\sqQVM{}}
\label{subsec:tilechain_sqQVM}
The \sqQVM{} shown in \cref{fig:TT_collection}(c) possesses $5$ distinct 't Hooft sectors $n_x = 0, 1, 2, 3$ and $4$, of which $n_x = 0$ and $n_x = 4$ possess a single state each with no dynamics. The sector $n_x = 1$ maps to the XXZ chain, similar to as discussed in \cref{subsec:XXZ_sqQVM}, and the sector $n_x = 3$ is equivalent to the sector $n_x = 1$ under the action of global spin flip $ \prod_l \sigma_l^x$.
The sector $n_x = 2$ is more interesting in that the configuration space is actually described in terms of $4$-tiles instead of $3$, with an associated Hilbert space dimension which scales with $\ell$ as $ ~ (1 + \sqrt{3})^{\ell}$ (see \cref{appendix:hilbertspacefragmentation}).

\begin{figure}
    \centerline{
    \includegraphics[height = 0.3\columnwidth]{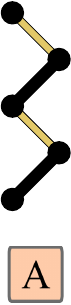} \includegraphics[height = 0.3\columnwidth]{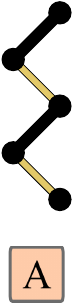} 
    ~~
    \includegraphics[height = 0.3\columnwidth]{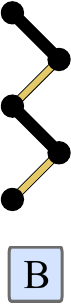}
    \includegraphics[height = 0.3\columnwidth]{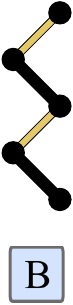}
    ~~~
    \includegraphics[height = 0.3\columnwidth]{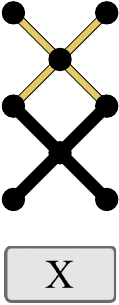}
    \includegraphics[height = 0.3\columnwidth]{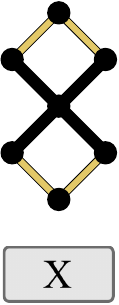}
    ~~~
    \includegraphics[height = 0.3\columnwidth]{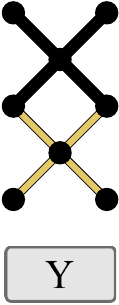}
    \includegraphics[height = 0.3\columnwidth]{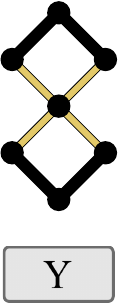}
    }
\caption{Explicit dictionary for mapping segments of the \sqQVM{} to tiles $\tileA$, $\tileB$, $\tileX$ and $\tileY$.}
\label{fig:sqQVM_dict}

\end{figure}

Under the explicit identification 
prescribed in \cref{fig:sqQVM_dict}, arbitrary string configurations in the $n_x = 2$ sector can be mapped to arrangements of tiles $\tileA$,~ $\tileB$,~ $\tileX$,~ $\tileY$ ~ (of lengths 1, 1, 2, and 2 respectively) into a sequence of length $2 \ell$. The hamiltonian of the model in the 4-tile Hilbert space is described in terms of RK kinetic terms
\begin{equation}
\begin{split}
    \hat{K}_I &= \ket{\hspace{0.004\columnwidth}\tileX} \bra{\tileA \tileB}_{I, I+1} \\
    &+ \ket{\hspace{0.004\columnwidth}\tileX} \bra{\tileB \tileA}_{I, I+1} \\
    &+ \ket{\hspace{0.004\columnwidth}\tileY} \bra{\tileA \tileB}_{I, I+1} \\
    &+ \ket{\hspace{0.004\columnwidth}\tileY} \bra{\tileB \tileA}_{I, I+1} \\
    &+ h.c.,
\end{split}
\end{equation}
and potential terms
\begin{equation}
\begin{split}
    \hat{P}_I &= \ket{\hspace{0.004\columnwidth}\tileA \tileB} \bra{\tileA \tileB}_{I, I+1} \\
    &+ \ket{\hspace{0.004\columnwidth}\tileB \tileA} \bra{\tileB \tileA}_{I, I+1} \\
    &+ \ket{\hspace{0.004\columnwidth}\tileX}\bra{\tileX}_{I, I+1} \\
    &+ \ket{\hspace{0.004\columnwidth}\tileY}\bra{\tileY}_{I, I+1},
\end{split}
\end{equation}

as
\begin{equation}
    \hat{H} = \sum_{I = 1}^{\ell_x - 1} -  t ~ \hat{K}_I + 2V \hat{P}_I.
\end{equation}

Instead of using $\tileX$ and $\tileY$ to describe the Hilbert space, one can perform a change of basis to their symmetric and antisymmetric combinations
\begin{align}
    \ket{\hspace{0.004\columnwidth}\tileS}_{I, I+1} &= \frac{\ket{\hspace{0.004\columnwidth}\tileX}_{I, I+1} + \ket{\hspace{0.004\columnwidth}\tileY}_{I, I+1}}{\sqrt 2}, \\
    \ket{\hspace{0.004\columnwidth}\tileAS}_{I, I+1} &= \frac{\ket{\hspace{0.004\columnwidth}\tileX}_{I, I+1} - \ket{\hspace{0.004\columnwidth}\tileY}_{I, I+1}}{\sqrt 2},
\end{align}
which has the advantage that kinetic terms simplify to
\begin{equation}
\begin{split}
    \hat{K}_I &= \sqrt{2} \ket{\hspace{0.004\columnwidth}\tileS}_{I, I+1} \bra{\tileA \tileB}_{I, I+1} + h.c. \\
    &+ \sqrt{2} \ket{\hspace{0.004\columnwidth}\tileS}_{I, I+1} \bra{\tileB \tileA}_{I, I+1} + h.c. ~ . \\
\end{split}
\end{equation}
Since $\tileAS$ does not explicitly appear in any of the kinetic terms and commutes with the potential terms, this tile has no dynamics. Equivalently, the projector
\begin{equation}
    \hat{\Pi}_I = \ket{\hspace{0.004\columnwidth}\tileAS}_{I,I+1}\bra{\tileAS}_{I,I+1}
\end{equation}
is a conserved quantity, i.e. it commutes with the hamiltonian $[ \hat{H}, \hat{\Pi}_I] = 0$. The presence of an extensive number of conserved quantities divides the Hilbert space into an exponentially large subspaces with dimensionality $\sim \phi^\ell$ where $\phi$ is the golden ratio (see \cref{appendix:hilbertspacefragmentation}), which can be viewed as a kind of local Hilbert space fragmentation \cite{Moudgalya_2022}.

The largest fragment of the Hilbert space is one in which there are no $\tileAS$ tiles, and this fragment maps exactly to the three-tile chain model from \cref{eqn:TileModel} with parameters $t' = \sqrt{2} t$, ${\VS}' = {\VAB}' =  2 V$. One can argue that this fragment contains the global ground state of the model, using a variational argument discussed in \cref{appendix:tilechain_groundstatefragment}.

\begin{figure}
    \centering
    \includegraphics[width = 0.83\columnwidth]{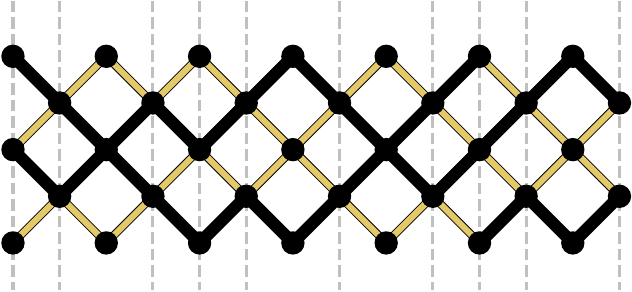}
    \includegraphics[width = 0.8\columnwidth]{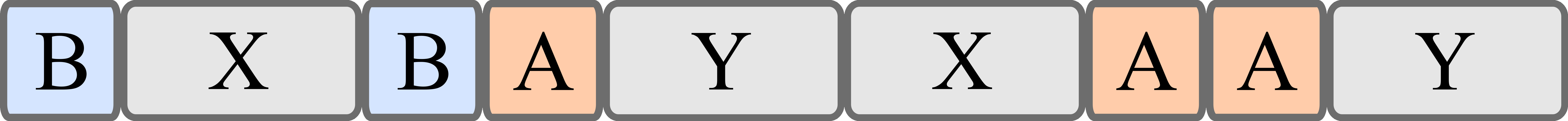}

    \caption{Identification of configurations in the two-string sector of \sqQVM{} with tilings spanned by $\tileA$, $\tileB$, $\tileX$ and $\tileY$.}
    \label{fig:TT_QVM}
\end{figure}

\subsection{Discussion}

\subsubsection{Symmetries of the model}
\label{subsubsec:tilechain_symmetries}
The model from \cref{eqn:TileModel} possesses a global $\mathbb{Z}_2$ flip symmetry implemented by the operator $\hat{F}$, defined via
\begin{equation}
\begin{split}
    \hat{F} \ket{\hspace{0.004\columnwidth}\tileA}_I &= \ket{\hspace{0.004\columnwidth}\tileB}_I, \\
    \hat{F} \ket{\hspace{0.004\columnwidth}\tileB}_I &= \ket{\hspace{0.004\columnwidth}\tileA}_I, \\
    \hat{F} \ket{\hspace{0.004\columnwidth}\tileS}_{I, I + 1} &= \ket{\hspace{0.004\columnwidth}\tileS}_{I, I + 1}. 
\end{split}
\end{equation}
In addition, the model possesses translation symmetry $\mathbb{Z}_l$, which shifts the positions of tiles by one unit length 
\begin{equation}\label{Tsymm}
\begin{split}
    \hat{T} \ket{\hspace{0.004\columnwidth}\tileA}_I &= \ket{\hspace{0.004\columnwidth}\tileA}_{I+1}, \\
    \hat{T} \ket{\hspace{0.004\columnwidth}\tileB}_I &= \ket{\hspace{0.004\columnwidth}\tileB}_{I+1}, \\
    \hat{T} \ket{\hspace{0.004\columnwidth}\tileS}_{I, I + 1} &= \ket{\hspace{0.004\columnwidth}\tileS}_{I+1, I + 2}.
\end{split}
\end{equation}
The $U(1)$ gauge structure of the original model survives in the tile chain as a $U(1)$ global symmetry $e^{i \mathcal{N} \theta}$ where the global charge is defined to be
\begin{equation}
\label{eqn:tilechain_charge}
\begin{split}
    \mathcal{N} = \sum_I  \ket{\hspace{0.004\columnwidth}\tileA} \bra{\tileA} _I -  \ket{\hspace{0.004\columnwidth}\tileB} \bra{\tileB} _I.
\end{split}
\end{equation}
All three of the symmetry operations commute, which implies that they combine into the symmetry group $\mathbb{Z}_2 \times \mathbb{Z}_\ell \times U(1)$.

\subsubsection{Long-range orders}
We find that the gapped ground states of the model \cref{eqn:TileModel} can be characterized by the symmetries that they spontaneously break. The numerical evidence we will present suggests that translation by two units (action of $\hat{T}^2$) is always a symmetry of the ground state, while $\hat{T}$ itself may be broken. 
There are three symmetry-breaking orders found for the model, which owing to their similarities to orders found in spin-1/2 chains, are called the \textbf{ferromagnetic} (FM), \textbf{antiferromagnetic} (AFM) and \textbf{valence bond solid} (VBS) orders, and are characterized as follows:
\begin{itemize}
    \item \textbf{Ferromagnetic phase} corresponds to an unbroken translation symmetry $\hat{T} = 1$, while the flip symmetry $\hat{F}$ is spontaneously broken. This phase is characterized by the two-fold degenerate ground states, similar to
    \begin{equation*}\label{Fsymm}
    \begin{split}
    \cdots ~ \tileA \tileA \tileA \tileA \tileA \tileA \tileA \tileA ~ \cdots~, \\
    \cdots ~ \tileB \tileB \tileB \tileB \tileB \tileB \tileB \tileB ~ \cdots ~. 
    \end{split}
    \end{equation*}

    \item \textbf{Valence bond solid phase} corresponds to an unbroken flip symmetry $\hat{F} = 1$, while the translation symmetry is spontaneously broken down to $\hat{T}^2 = 1$. This phase is characterized by two-fold degenerate ground states exemplified by
    \begin{equation*}
        \begin{split}
             ~ \cdots ~ \tileS \tileS \tileS \tileS ~ \cdots ~ ,   \hspace{0.5cm} \\
            \hspace{0.5cm} ~ \cdots ~  \tileS \tileS \tileS \tileS ~ \cdots ~ .
        \end{split}
    \end{equation*}

\item \textbf{Antiferromagnetic phase} corresponds to an unbroken `translate-flip' symmetry $\hat{F} \hat{T}$, while the individual flip $\hat{F}$ and translation $\hat{T}$ symmetries are both broken spontaneously. This is similar to Neel-order in spin chains, exemplified by
\begin{equation}
    \begin{split}
    ~ \cdots ~ \tileA \tileB \tileA \tileB \tileA \tileB \tileA \tileB ~ \cdots ~ , \\
    ~ \cdots ~ \tileB \tileA \tileB \tileA \tileB \tileA \tileB \tileA ~ \cdots ~ .
    \end{split}
    \end{equation}
\end{itemize}

\begin{figure}
    \centering
    \includegraphics[width=\columnwidth]{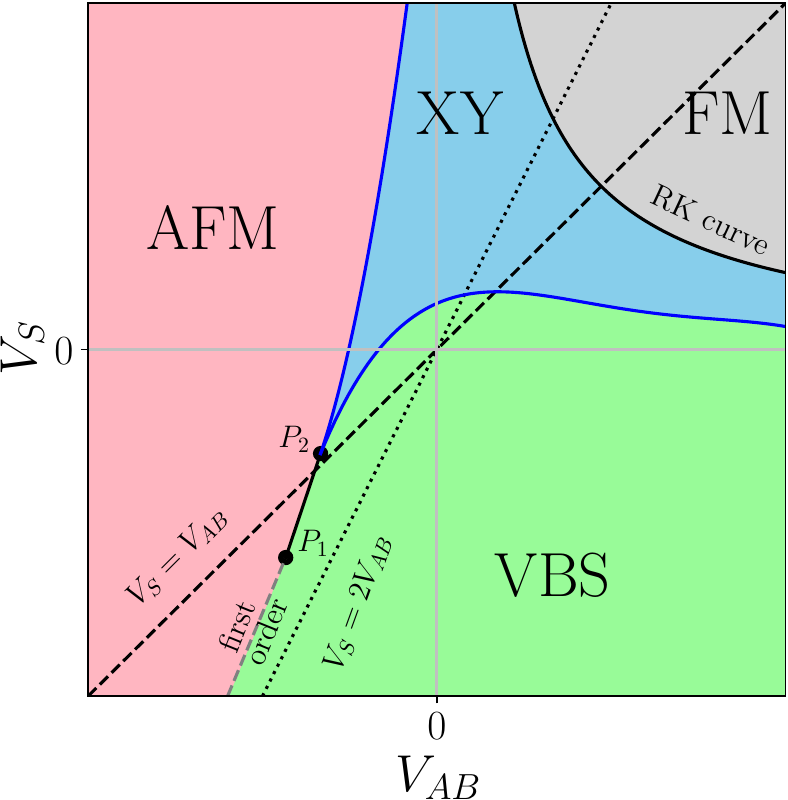}
    \caption{Schematic phase diagram for the tile chain. The lines $\VAB = \VS$ and $\VS = 2\VAB$ correspond to the models \sqQVM{} (see \cref{subsec:tilechain_sqQVM}) and \trQVM{} (see \cref{subsec:tiles_QVMtriangular}) respectively. Based on DMRG simulations on moderately sized systems, we find reasonable evidence that the line $\VS = \VAB$ intersects the Landau-forbidden line of critical points $P_1 - P_2$.}
    \label{fig:Tile_schematic_dqcp}
\end{figure}

\subsubsection{RK curve}
The ground state of the tile chain can be exactly solved for a continuous curve of points in parameter space,
\begin{equation}
 \VAB \VS = 2 t^2 ~~~~ (\VAB > 0 ),
\end{equation}
where the hamiltonian reduces to a positive sum of projectors analogous to the RK point in the original models.
\begin{equation}
\begin{split}
    \frac{H}{\VAB} = \sum_I &  \left( \ket{\hspace{0.004\columnwidth}\tileA \tileB} - \frac{t}{\VAB} \ket{\hspace{0.004\columnwidth}\tileS} \right) \left( \bra{\tileA \tileB} - \frac{t}{\VAB} \bra{\tileS} \right) \\
     + &  \left( \ket{\hspace{0.004\columnwidth}\tileB \tileA} - \frac{t}{\VAB} \ket{\hspace{0.004\columnwidth}\tileS} \right) \left( \bra{\tileB \tileA} - \frac{t}{\VAB} \bra{\tileS} \right)
\end{split}
\end{equation}
The ground state can be constructed as the weighted sum of all allowed configurations
\begin{equation}
    \ket{RK} = \sum_{ \textrm{all } \psi} \left( \frac{t}{\VAB}\right)^{- N_S(\psi)} \ket{\psi},
\end{equation}
where $\ket{\psi}$ is a ket belonging to the configuration basis, and $N_S(\psi)$ is the number of $\tileS$ in the configuration associated with $\psi$. By construction, this state has energy $0$ and lies in the simultaneous $0$-eigenspace of the projectors. As written, the RK wavefunction is not an eigenstate of the $U(1)$ charge of the model $\mathcal{N}$ defined in \cref{eqn:tilechain_charge}, but a linear combination of ground states lying in several $U(1)$ sectors.

\subsubsection{Classical limit $t \rightarrow 0$}
The limit where $|\VAB|, |\VS| \gg |t|$ can be understood in terms of the corresponding classical problem, where the configuration with the least (potential) energy is  the ground state. Parametrizing the classical model in terms of a single angle $\theta$ as $\VAB / \cos \theta = V_S / \sin \theta = \sqrt{\VAB^2 + V_S^2}$, we find 

\begin{equation}
\begin{split}
    0 < ~  & \theta < \pi/2 \hspace{2.2cm} \textrm{ FM} \\
    \pi/2 < ~ & \theta < \pi + \tan^{-1}(2) \hspace{0.8cm} \textrm{AFM} \\
    \pi + \tan^{-1}(2) < ~ & \theta < 2 \pi \hspace{2.4cm} \textrm{VBS} 
\end{split}
\end{equation}

The points $\theta = 0,~ \pi/2 $ and $\pi + \tan^{-1}(2)$ correspond to points where the different gapped long-range orders show a transition under the assumption $t = 0$. The above lines define the phase transitions in the phase diagram from Fig.\ref{fig:Tile_schematic_dqcp} asymptotically away from the origin. In some cases, we can uncover the nature of transitions by considering the effect of $t \ll \VAB, \VS$ perturbatively.

\subsubsection{Transition near $\theta = \pi/2$}
Around $\theta = \pi/2$, $\tileS$ are strongly suppressed by the energy scale $V_S$, and one can consider an effective model over the space of tilings generated by $\tileA$, $\tileB$. This subspace is isomorphic to that of a spin-1/2 chain of length $\ell$. To the lowest order in $t$, one finds that the effective dynamics of the model is governed by 
\begin{equation}
\label{eqn:tiles_XXZlimit}
\begin{split}
    \hat{K}_{\textrm{effective}} &= - \frac{t^2}{V_S} \sum_{I} \ket{\tileA \tileB} \bra{\tileB \tileA} + h.c. ~, \\
    \hat{V}_{\textrm{effective}} &= \left( \VAB - \frac{t^2}{V_S} \right) \sum_{I} \ket{\tileA \tileB} \bra{\tileA \tileB} \\
    &+ \left( \VAB - \frac{t^2}{V_S} \right) \sum_{I} \ket{\tileB \tileA} \bra{\tileB \tileA}.
\end{split}
\end{equation}
where the $t^2/\VS$ terms arise when perturbatively considering the effect of the subspace with at least one $\tileS$ in the configuration. Thus, to the lowest order in $t/V_S$, the model is described by an XXZ chain with anisotropy $v = V_S \VAB/t^2 - 1$, suggesting a KT phase transition for the model close to the line $\VAB = 0$ in the limit $\VS \rightarrow  + \infty$. Furthermore, it is consistent with first order transition expected at the RK curve $\VAB = 2t^2/V_S$.

\subsubsection{Transition near $\theta = \pi + \tan^{-1} 2$}
In the negative cuadrant ($V_{S}<0$ and $V_{AB}<0$) and along the line  $V_{S} = 2 V_{AB}$, the AFM and VBS states are degenerate in the classical limit ($t/V_{AB} \rightarrow 0$), exemplified by
\begin{equation}
\begin{split}
   \cdots ~ \tileA \tileB \tileA \tileB \tileA \tileB \tileA \tileB \tileA \tileB ~ \cdots, \\
    \cdots ~ \tileS \tileS \tileS \tileS \tileS ~ \cdots.
\end{split}
\end{equation}
We can argue that in the classical limit there is a first order transition between the two phases, because the energy of a single domain wall between the two phases is of the order $\sim \VAB/2 + O(t^2/\VAB)$, where $t^2/\VAB$ term is a perturbative correction. 
\begin{equation}
\begin{split}
    \cdots ~ \tileA \tileB \tileA \tileB \tileA \tileB \tileS \tileS \tileS ~ \cdots,\\
    \cdots ~ \tileA \tileB \tileA \tileB \tileS \tileS \tileS \tileS ~ \cdots,
\end{split}
\end{equation}
Thus, for $t \ll \VAB$, one can bound the energy of each domain wall to be of order $\VAB$, implying that they cannot proliferate to give rise to a intermediate gapless disordered phase.

\subsubsection{Mapping to constrained fermion model}\label{hardcore-Rep}

\begin{figure}
    \centering
    \includegraphics[height=0.3\columnwidth]{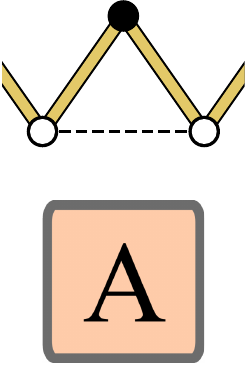} ~ \includegraphics[height=0.3\columnwidth]{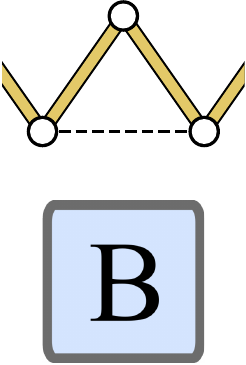} ~ \includegraphics[height=0.3\columnwidth]{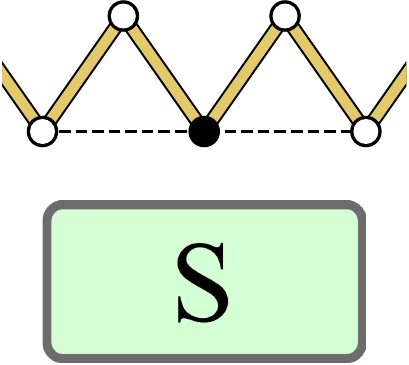}
    \caption{Mapping between the tile picture and the hardcore boson chain.}
    \label{fig:hardcorechain_dictionary}
\end{figure}

\begin{figure}
    \centering
    \includegraphics[width=0.9 \columnwidth]{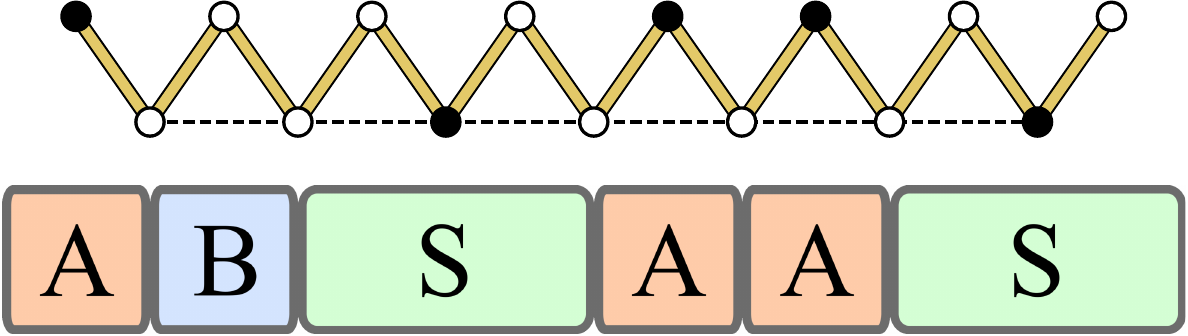}
    \caption{Demonstration of the mapping between the tile chain and the hardcore chain. The upper row corresponds to the even sites while the lower row corresponds to the odd sites.}
    \label{fig:hardcorechain_demo}
\end{figure}

\begin{figure}
    \centering
    \includegraphics[width=0.9\linewidth]{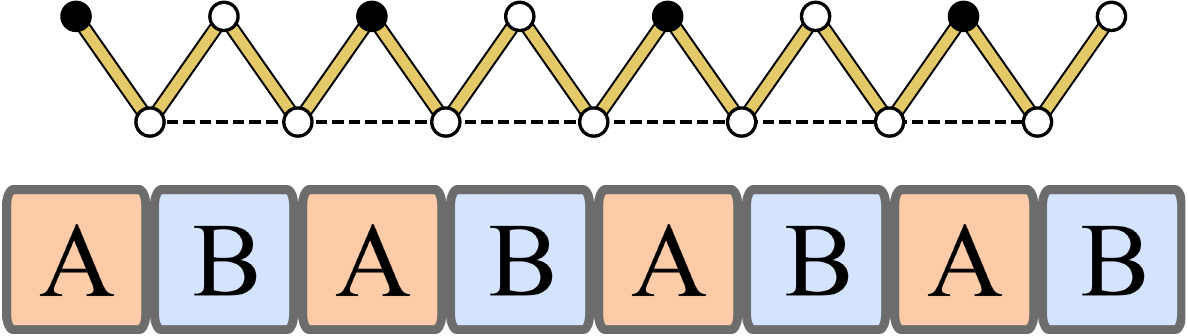} \\ ~ \\ ~ \\
    \includegraphics[width=0.9\linewidth]{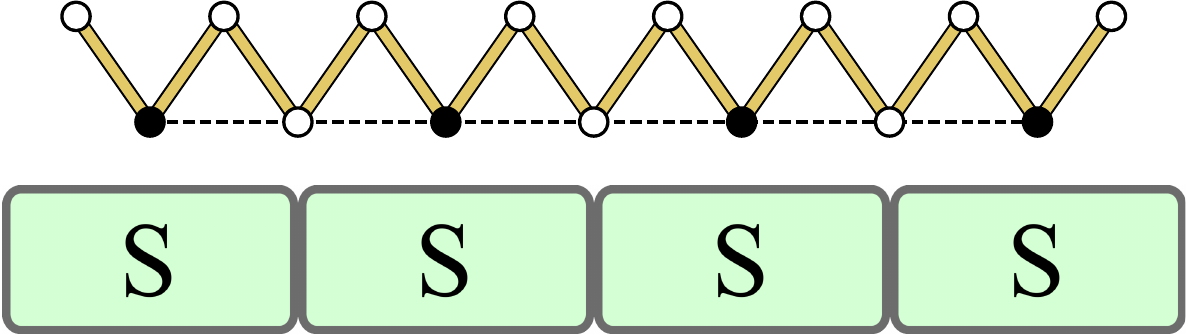}
    \caption{The VBS and the AFM phases can be interpreted as $1/4$-filled charge density waves in the chain picture.}
    \label{fig:hardcorechain_ABS}
\end{figure}

The configurations of three-tile model of length $\ell$ can also be mapped onto a model of hardcore bosons on a chain with $2 \ell$ sites (or equivalently onto spinless fermions via Jordan-Wigner transformation). However not all boson configurations are physical configurations of tiles, and therefore some of them need to be eliminated by imposing local constraints. More specifically, the tile Hamiltonian from Eq.\eqref{eqn:TileModel} maps onto the following Hamiltonian of hard-core bosons:
\begin{align*}
H &= -t \sum_{i} b^\dagger_{i+1} b^{}_{i} + b^\dagger_{i} b^{}_{i+1} \\ 
    &+ V_{\infty} \sum_{i}  n_{i} n_{i + 1} \\
    &+ V_{\infty} \sum_{even ~ i} n_{i} n_{i + 2} \\
    &+ \VS \sum_{odd ~ i} n_{i} \\
    & - \VAB \sum_{even ~ i} n_{i} n_{i + 2} \\
    & + 2 \VAB \sum_{even ~ i} n_{i} \\
    & - \VAB \sum_{i} n_{i} n_{i + 3},
\end{align*}
where the potential $V_{\infty}$ is formally taken to be $\infty$ to enforce constraints in the model. This mapping is obtained by performing the identifications shown in the \cref{fig:hardcorechain_dictionary}, where the upper sites are those with odd $i$ and the lower sites are those with even $i$. A dotted line has been drawn between even-numbered sites to emphasize the hardcore repulsion between them. An explicit example for this mapping for an arbitrary state has been shown in \cref{fig:hardcorechain_demo}.

Despite being redundant, this representation has the advantage that is easier to simulate numerically with open boundary conditions using existing DMRG packages, such as ITensor \cite{ITensor_Stoudenmire}. This is done by imposing the hardcore interaction by means of a projector \cite{Projector_Dalmonte,RydbergLadder_LuisaEck}. Moreover, this representation allows the interpretation of both the AFM and VBS phases of the tile chain as charge density waves in the hardcore bosonic representation, which in turn facilitates the usage of CDW order parameters to diagnose the phases numerically. Another advantage of this representation is that it will allow us to understand phase transitions via a field-theoretic bosonization \cite{Giamarchi} of the hardcore boson chain, which will be particularly advantageous to understand the possibility of the deconfined critical point that we will discuss in the next section.

However, some symmetries of the tile representation become less transparent in terms of hard-core bosons. In particular, the $\mathbb{Z}_2$ symmetry of the tile chain denoted by $\hat{F}$  (see Eq.\eqref{Fsymm}) is implemented in the hardcore boson model unusually in that it is only a symmetry in the presence of hardcore repulsive terms $V_{\infty} = \infty$, and does not survive if the repulsion is softened. This symmetry can be represented as a sum of local terms of the PXP form \cite{Original_PXPmodel}:
\begin{equation}
    \hat{F} = \sum_{\textrm{odd} ~ i} (1 - n_{i - 1}) (b^\dagger_i + b_i) (1 - n_{i + 1}).
\end{equation}

Interestingly the above symmetry forces the hard-core bosons to be at a lattice filling $1/4$. We also note that the translational symmetry $\hat{T}$ from Eq.\eqref{Tsymm} is implemented as a translations by two sites $i\rightarrow i+2$ in the hard-core bosonic representation, since the number of sites has been doubled (see e.g. Fig.\ref{fig:hardcorechain_ABS}).

\subsubsection{Deconfined quantum criticality}
\begin{figure}
    \centering
    \includegraphics[width=\columnwidth]{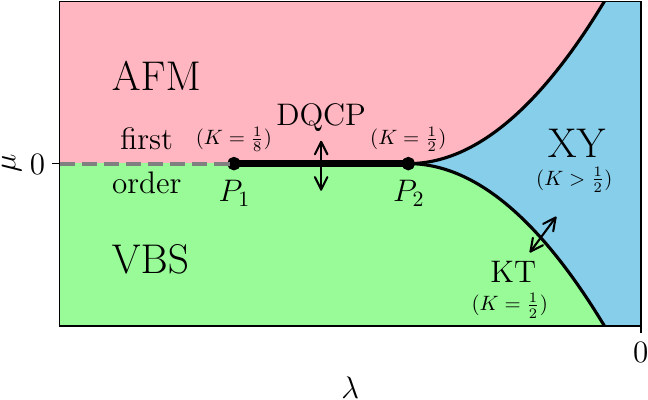}
    \caption{Schematic of Landau-forbidden quantum critical points with details of the Luttinger parameter.}
    \label{fig:schematic_dqcp}
\end{figure}

Our previous analysis and numerical findings indicate that except in the FM region above the RK line (see Fig.\ref{fig:Tile_schematic_dqcp}), either $\hat{F}$ or $\hat{F}\hat{T}$ remains a symmetry of the ground state, and therefore the hard-core bosons remain at filling $1/4$ of the lattice throughtout the AFM, VBS, and XY phases, as well the phase transitions between them. On the other hand, the AFM and VBS phases are characterized by unbroken symmetries $\hat{F} \hat{T}$ and $\hat{F}$ respectively, which implies that they correspond to symmetry-breaking orders to distinct subgroups. Therefore, they break distinct microscopic symmetries, and according to the Landau theory of phase transitions \cite{LandauTheory}, there cannot be a continuous phase transition between such phases without fine-tuning. However, it has been proposed that such critical points are possible, and can be referred to as deconfined quantum critical points (DQCP) \cite{OriginalDQCP_Senthil,DQCP_review}. 

We will provide evidence to support that the tile chain shows a line of Landau-forbidden critical points in the $\VAB - \VS$ parameter space, which can be understood via an effective field theory description expected to be valid in the gapless phases,
\begin{equation}
\label{eqn:dqcp_action}
\begin{split}
    \mathcal{S} = \frac{1}{4 \pi K} \int dx ~ d\tau ~ & u ( \partial_x \phi )^2 + u^{-1} ( \partial_\tau \phi )^2 \\
    &+ \mu \cos( 4 \phi) + \lambda \cos( 8 \phi) + \cdots.
\end{split}
\end{equation}
Such a field theory description is analogous to that of a (bosonized) Luttinger liquid, where the cosine terms, when relevant, can drive the system to one of four, $1/4$-filled commensurate charge density wave (CDW) phases. These phases correspond to $\phi = 0, \pi/2$ (AFM) and $\phi = \pi/4, 3\pi/4$ (VBS), with respective order parameters $\cos(2 \phi)$ and $\sin(2 \phi)$ respectively. In fact, the tile chain can be mapped to a fermionic chain with hardcore repulsion terms which shows such CDW phases and suggests the description to be valid (see \cref{sec:hardcore_model} for more details).

We argue for the existence of a line of Landau-forbidden critical points by considering the scaling dimensions of $\mu$ and $\lambda$ depending on the Luttinger parameter $K$. It can be shown that for $K > 1/2$, both $\mu$ and $\lambda$ are irrelevant, which we claim corresponds to the stable Luttinger liquid phase found in the model. For $K < 1/2$, the $\mu$ term is relevant, causing the Luttinger liquid to generically gap out into the AFM or VBS phases whenever $\mu \neq 0$. An exception occurs along the line $P_1 - P_2$, where the parameter $\mu$ is $0$, by virtue of lying between regions with $\mu > 0$ and $\mu < 0$. Restricted to this line, $\lambda$ is the coefficient of the leading cosine term, and is irrelevant when $K > 1/8$. Since we generically expect (barring fine-tuned cases) the Luttinger parameter $K$ to vary continuously as the function of microscopic parameters $\VAB, \VS$, this suggests that there is a continuous line $P_1 - P_2$ (see \cref{fig:schematic_dqcp}) of quantum critical points between AFM and VBS phases. Along this line, $K$ varies from $1/2$, at the multi-critical point $P_2$, to $1/8$, at $P_1$, the boundary of the first-order transition line. Our treatment is similar to that of Landau-forbidden criticality explored in other quantum chains \cite{NeelVBS_Haldane,BeyondLandau_Mudry,LandauForbidden_SoonwonChoi}. In \cref{sec:hardcore_model} we provide further concrete numerical evidence for the prediction $K \in (1/8, 1/2)$ from the existence of a DQCP in the \sqQVM{}.

\subsubsection{$v_{AB} =  2 v_{S} = 2v$}
\begin{figure}
    \centering
    \includegraphics[width=\columnwidth]{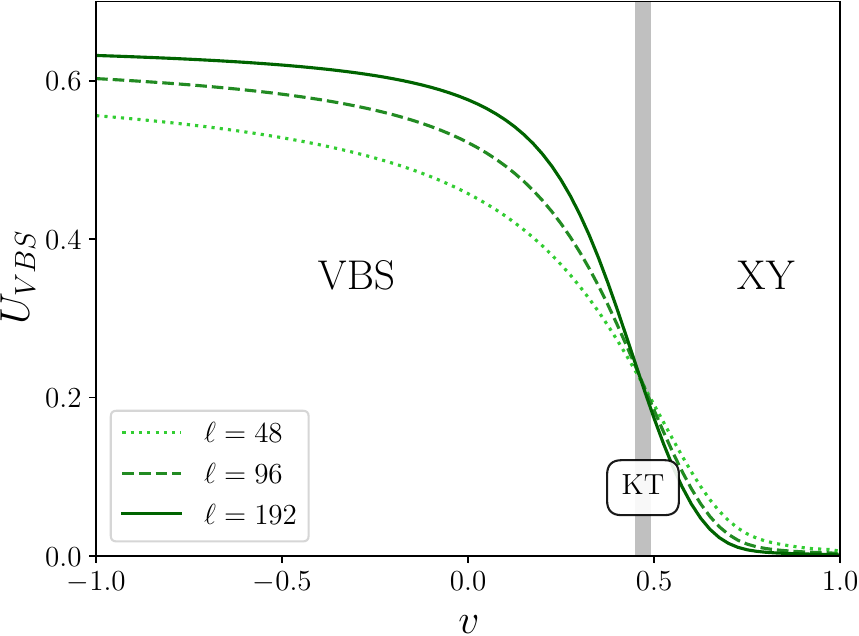} \\ ~ \\
    \includegraphics[width=\columnwidth]{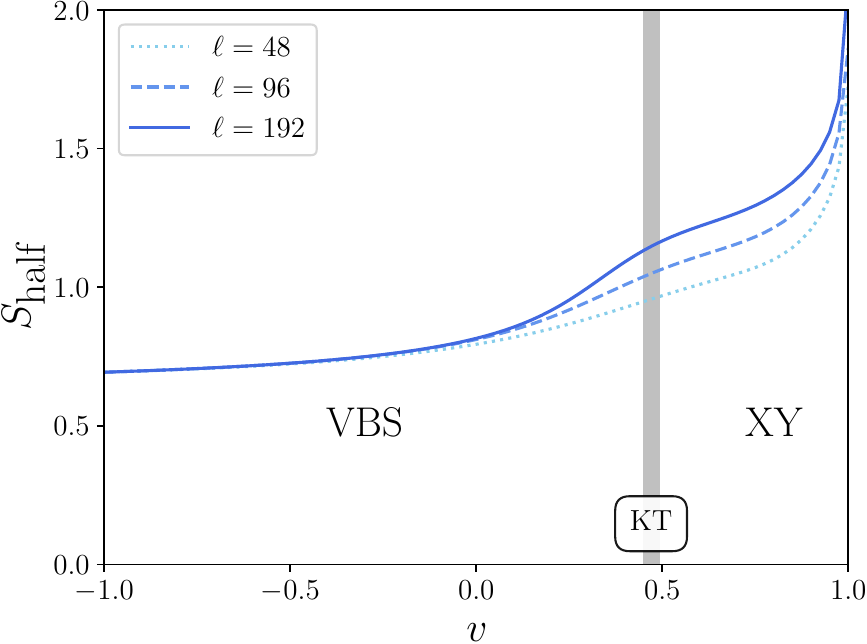}
    
    \caption{\textbf{Top:} Binder cumulant associated with the order parameter for the VBS phase in the system. \textbf{Bottom:} Half-cut entanglement entropy.}
    \label{fig:Tile1_VBS}
\end{figure}

This model, which corresponds to the dotted line in Fig.\ref{fig:Tile_schematic_dqcp}, has been studied before in Ref.\cite{JonahSebaInti}, where the authors concluded a very wide domain of gaplessness based on exact-diagonalization calculations in small systems. We attribute this seemingly extended region to the proximity to first order and continuous DQCP phase transitions into the AFM phase, as shown in \cref{fig:Tile_schematic_dqcp}. Proximity to such transitions causes correlation lengths to be large, which may be mistaken for gaplessness for small system sizes. In our DMRG analysis (see \cref{fig:Tile1_VBS}) we find that the gapped VBS phase persists up to $v_{KT} = 0.47 \pm 0.03$, at which point it shows a Kosterlitz-Thouless transition into the XY (or Luttinger liquid) phase, which persists until the RK point $v = 1$.

\begin{figure}
    \centering
    \includegraphics[width = 0.4\columnwidth]{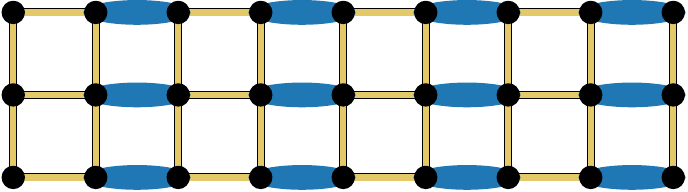} ~~
    \includegraphics[width = 0.4\columnwidth]{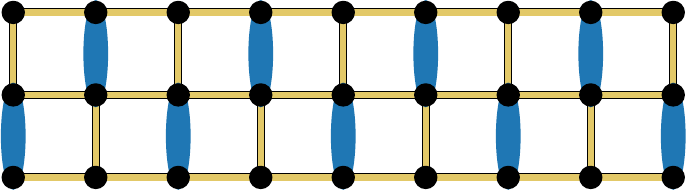} \\ ~ \\

    \includegraphics[width = 0.8 \columnwidth]{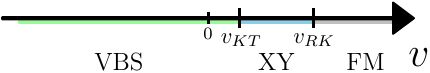}
    \caption{Phase diagram for the quantum dimer ladder admitting a single string description. Representative states have been chosen for each solid phase in the lattice gauge theory.}
    \label{fig:pd_OneString}
\end{figure}

\subsubsection{$v_{AB} =  v_{S} = \sqrt{2} v$}
\begin{figure}
    \centering
    \includegraphics[width=\columnwidth]{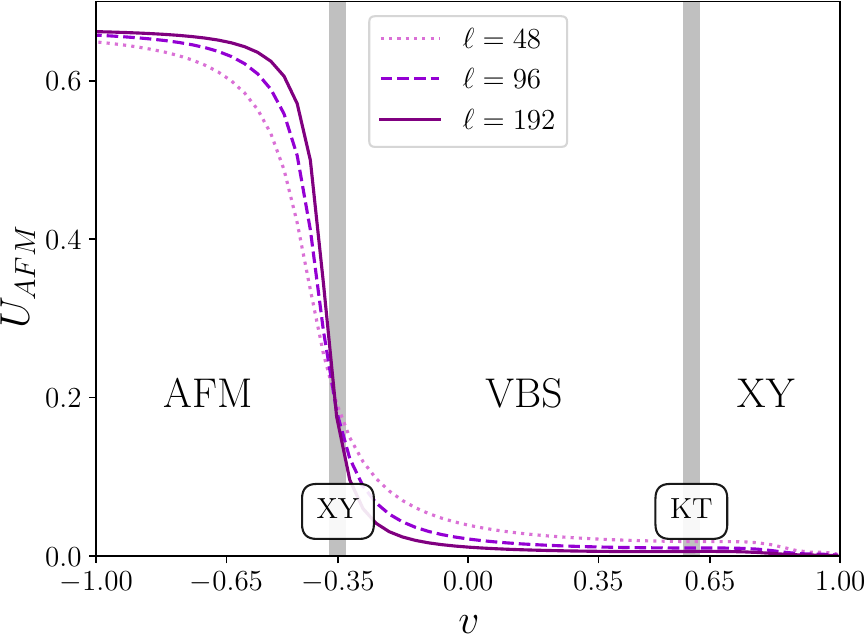} \\ ~ \\
    \includegraphics[width=\columnwidth]{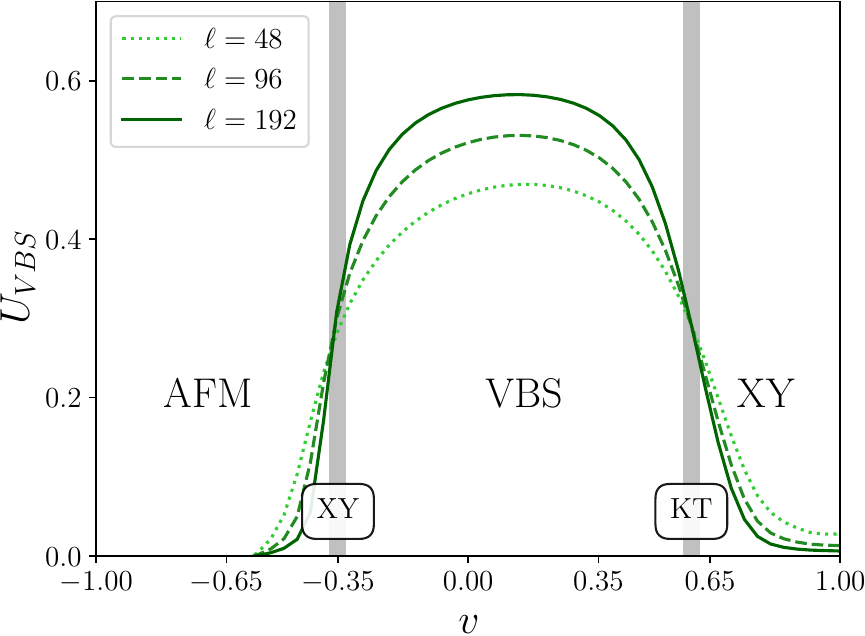}  \\ ~ \\
    \includegraphics[width=\columnwidth]{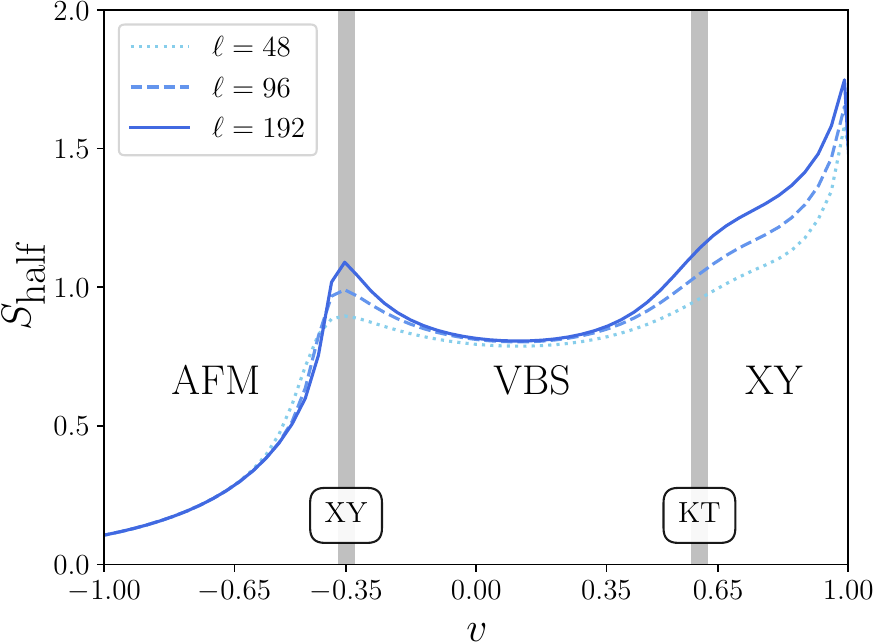}
    
    \caption{\textbf{Top:} Binder cumulant associated with the AFM order parameter. \textit{Middle:} Binder cumulant associated with the VBS order parameter. \textit{Bottom:} Half cut entanglement entropy. }

    \label{fig:Tile2_Numerics}
\end{figure}

\begin{figure}
    \centering
    \includegraphics[width = 0.3\columnwidth]{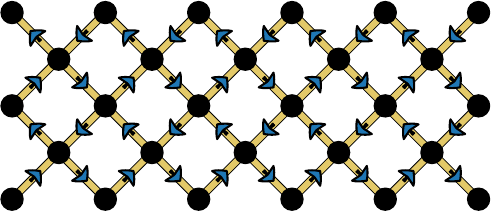} \hspace{0.0\columnwidth}
    \includegraphics[width = 0.3\columnwidth]{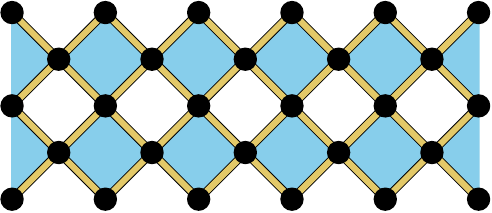} 
    \includegraphics[width = 0.3\columnwidth]{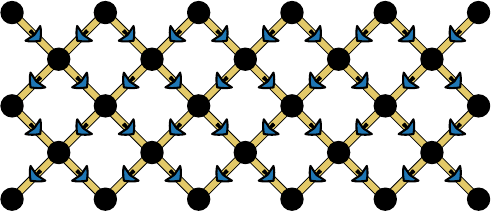} \\ ~ \\

    \includegraphics[width = \columnwidth]{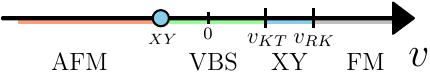}
    \caption{Phase diagram for the \sqQVM{} admitting a two-string description. Representative states have been chosen for each solid phase in the lattice gauge theory.}
    \label{fig:pd_SpinOne_half}
\end{figure}

This model which corresponds to the dashed line in Fig.\ref{fig:Tile_schematic_dqcp}, and that has the phase diagram shown in \cref{fig:pd_SpinOne_half}, and its behavior can be summarized as follows:
\begin{enumerate}
    \item $v < v_{XY}$: (Gapped, GSD = 1). AFM phase. \\
    \item $v = v_{XY}$: (Gapless, $\omega \sim k$). Believed to be a deconfined quantum critical point, $c = 1$.
    \item $ v_{XY} < v < v_{KT} $: (Gapped, $\omega \sim k$). VBS phase, quasi-long-range ordered descendant of the resonating plaquette solid in the full 2D problem.
    \item $ v = v_{KT} \approx $: (Gapless, $\omega \sim k$). Location of KT phase transition into the gapless luttinger liquid phase.
    \item $v_{KT} < v < v_{RK}$: (Gapless, $\omega \sim k$). Gapless phase.
    \item $v = v_{RK} = 1$: (Gapless, $\omega \sim k^2$). RK point, first-order transition into the ferromagnetic phase.
    \item $v > v_{RK} = 1$: (Gapped, GSD = 2). $\bigotimes_I \ket{\tileA}_I$ and $\bigotimes_I \ket{\tileB}_I$ are the exact ground states here.
\end{enumerate}
Numerically, we find the DQCP to be at $v_{XY} = - 0.35 +/- 0.02$, and $v_{KT} = 0.60 +/- 0.02$, as shown in \cref{fig:Tile2_Numerics}. Binder cumulants for the AFM and VBS phases have been defined in \cref{sec:hardcore_model}.

\section{Summary and Discussion}
\label{sec:discussion}
We have studied a large class of $U(1)$ lattice gauge theories described by Rokhsar-Kivelson models compactified on narrow cylinders, which map exactly onto three 1D spin chain models: (i) the XXZ chain, (ii) a spin-1 chain, and (iii) a chain with extra constraints that is more efficiently described as one-dimensional model of tiles. These mappings were developed by analyzing the string representations of these models, which follow from the $U(1)$ gauge structure and topological conservation laws of the model. Interestingly, we have shown that the global transverse momentum of the string maps exactly onto the twist of boundary conditions for the 1D spin chain. This allowed us to show that the inverse mass of the global string motion is identical to the internal Drude weight of its corresponding spin chain. 

The spin-chains thus obtained show phase transitions into a fluid XY phase, which we interpret to be a quasi-long range ordered descendant of the resonant 2D plaquette phases. Indeed, we find the numerically obtained value of $v_{1D}$ at which the spin chain displays onset of fluid order, to be reasonably proximate to the values of $v_{2D}$ at which one finds onset of resonant plaquette order in the 2D problem. For example, in the quantum dimer model on the hexagonal lattice, the spin-1 chain predicts $v_{1D} = -0.4 \pm 0.1$, which is close to numerically obtained $v_{2D} \sim -0.2$ \cite{hxQDM_Moessner}. For the square lattice, it is unclear \cite{sqQDMnumerics_latest,MixedPhaseQDM_Moessner} whether a plaquette phase even exists in the 2D problem. Nevertheless, our finding onset of fluidity at $v_{1D} \sim 0.5$ is consistent with the quantum Monte Carlo studies \cite{QMC_QDMplaq_05,QMC_QDMplaq_06} which report the plaquette phase to exist for $v_{2D} > 0.6$. The tile chain predicts the transition between the two crystal phases of the quantum six-vertex model to be at $v_{1D} \sim -0.35$, which is remarkably close to the full 2D result $v_{2D} = -0.37$ \cite{sqQVMnumerics_Dalmonte} and $v_{2D} = -0.35$ \cite{sqQVM_numerics_rydberg}. Thus, despite their simplicity, this indicates the 1D models reasonably approximate the phase diagrams of their 2D counterparts.

Furthermore, we have uncovered many interesting new behaviors in this quasi-one-dimensional setting. For example, we found that at its RK point the spin-1 chain displays a continuum manifold of exact grounds states that is topologically equivalent to a  sphere, even though these states are distinct from spin-1 coherent states and there is no underlying microscopic $SU(2)$ symmetry. As a consequence, the stiffness of the gapless quadratically dispersing mode at its RK point changes continuously with the global magnetization of the spin-1 chain. In the case of the tile chain, we found that a continuum line of RK points where the relative density of different tile types changes continuously.

In addition to their success at describing the low energy physics, the process of mapping RK models to spin chains has allowed us to uncover emergent conservation laws. Since the XXZ model is known to be integrable via the Bethe ansatz, it immediately follows that all of the RK models described by the XXZ chain are integrable as well. Furthermore, we find local Hilbert space fragmentation in the quantum six-vertex ladder of length $\ell$, where the sector with dimension $\sim 2.73^\ell$ fractures into $\sim 1.62^\ell$ disconnected sectors, with the size of the largest sector growing as $\sim 2.41^\ell$, constituting a vanishingly small fraction in the limit of large $\ell$. The existence of such disconnected sectors is particularly interesting in the context of recent work \cite{Scars_RK_Banerjee_21,Scars_RK_Banerjee_22,Scars_Schwinger_Banerjee_2023,Scars_TruncatedSchwinger_Banerjee_2023,Scars_RK_Bannerjee_2024} on quantum many body scars in lattice gauge theories.

Finally, we also have uncovered evidence for the existence of a line of continuous phase transitions between gapped phases in the tile chain. Such a critical line is forbidden by the Landau theory of phase transitions, but we have argued it to exist in our model without fine-tuning. Our numerical investigation suggests that such a deconfined quantum critical point exists in the phase diagram for the quantum six-vertex ladder, in contrast to the standard $2D$ quantum six-vertex model \cite{QVMnoDQCP_Bannerjee}, which is fully gapped for all $v \neq 1$. Whether this construction suggests a way to access or realize deconfined or other gapless states in 2D RK-like models, such as those discussed in \cite{GollerInti}, is still an open problem, and another interesting avenue for future research.

Evidently, there is rich physics captured by effective spin-chain descriptions of RK models, ranging from low-energy properties, to Hilbert space structure and exotic criticality. Some of the future theory directions include applying the string framework for $U(1)$ lattice gauge theories other than ones described by RK hamiltonians, which may lead to richer examples of fragmentation and exotic quantum criticality. It will be interesting to try to understand confinement in gauge theories from the perspective of extended string objects introduced in \cite{JonahSebaInti} and this work.

\section*{Acknowledgements}
We are thankful for helpful discussions with Debasish Banerjee, Haruki Watanabe, Matthias Thamm, Johannes Stephan Hofmann, Abhishodh Prakash and Saranesh Prembabu. G.S. acknowledges financial support from the Working Internships in Science and Engineering (WISE) fellowship from the German Academic Exchange Service (Deutscher Akademischer Austauschdienst, DAAD) as well as the Kishore Vaigyanik Protsahan Yojana (KVPY) fellowship from the Dept. of Science and Technology, Government of India. We acknowledge support from the Deutsche Forschungsgemeinschaft (DFG) through research grant project numbers 542614019; 518372354; 555335098.

\clearpage

\appendix{}
\crefalias{section}{appendix}

\section{More details on hard-core boson representation of tile model and the DQCP}
\label{sec:hardcore_model}

\subsubsection*{Order parameters}
The order parameter for the AFM phase in the hardcore boson chain framework is
\begin{equation}
    M_{AB} = \frac{1}{\ell} \sum_{i = 0}^{2\ell - 1} n_i \cos \left( \frac{\pi}{2} i \right). 
\end{equation}
Similarly, for the VBS phase we have
\begin{equation}
    M_{S} = \frac{1}{\ell} \sum_{i = 0}^{2\ell - 1} n_i \sin \left( \frac{\pi}{2} i \right). 
\end{equation}
Accordingly, one can define Binder cumulants for these two order parameters
\begin{equation}
\begin{split}
    U_{AB} &= 1 - \frac{1}{3} \frac{\langle M^4_{AB} \rangle}{\langle M^2_{AB} \rangle^2}, \\
    U_{S} &= 1 - \frac{1}{3} \frac{\langle M^4_{S} \rangle}{\langle M^2_{S} \rangle^2}.
\end{split}
\end{equation}
We compute these numerically using DMRG. In addition, we look at the bipartite entanglement entropy as a proxy for fluidity in the system.

\subsubsection*{Tricritical point and DQCP}
Since the model is at $1/4$ - filling, we expect the effective (bosonized) field theory to have umklapp terms which may drive the transition to the charge density wave phases. Thus, we phenomenologically write down the field theory describing the critical phases
\begin{equation}
\begin{split}
    \mathcal{H} \propto \int dx ~ & \frac{1}{K} ( \partial_x \phi )^2 + K ( \partial_x \theta )^2 \\
    &+ \mu \cos( 4 \phi) + \lambda \cos( 8 \phi) + \cdots.
\end{split}
\end{equation}
Where the scalar fields $\phi, \theta$ satisfy the commutation relation $[\phi(x), \partial \theta(x')] = \ci \pi \delta(x - x')$, and are related to the hardcore-bosonic operators via the relations
\begin{equation}
    b^{\dagger}_{i} = \frac{e^{- \ci \theta(r_i)}}{\sqrt{2 \pi \alpha}} \left( A_1 \cos(i + \gamma_1) + A_2 \cos(2i + \gamma_2) + \cos(2 \phi(x)) \right) 
\end{equation}
where $A_1, A_2, \gamma_1, \gamma_2$ are constants sensitive to the microscopic details and
\begin{equation} 
n_{i} = \frac{1}{4} - \frac{1}{\pi} \nabla \phi(r_i) + \frac{1}{\pi \alpha}\begin{cases}
     + (-1)^{i/2} \cos(2 \phi(r_i)) & i \textrm{ even} \\
     + (-1)^{(i-1)/2} \sin(2 \phi(r_i)) & i \textrm{ odd}. \\
\end{cases}
\end{equation}
Here $\alpha$ is the UV regulator, and can be taken to be of the order of the lattice spacing $a$. The symmetries of the model include the global $U(1)$ symmetry, $\theta(x) \rightarrow \theta(x) + \alpha$, the flip symmetry $\phi(x) \rightarrow \pi - \phi(x)$. Under the identifications, we find that the order parameters of the AFM and VBS phases are given by $M_{AB} \propto \langle \cos(2 \phi) \rangle, ~ M_{S} \propto \langle \sin(2 \phi) \rangle.$ The density-density connected correlator on the even sites obeys 
\begin{equation}
    \langle n_{i} n_{i + r} \rangle - \langle n_{i} \rangle \langle n_{i + r} \rangle \sim C_1 \frac{1}{r^2} + C_2 \frac{(-1)^r}{r^{2K}},~~~ (i, r \textrm{ even}).
\end{equation}
Thus, the power law decay of the density-density correlator is governed by the critical exponent $\eta = 2K$, which can be probed numerically.

\subsubsection*{Numerical evidence for DQCP}
\label{subsubsec:numerical_DQCP}
\begin{figure}
    \centering
    \includegraphics[width=\columnwidth]{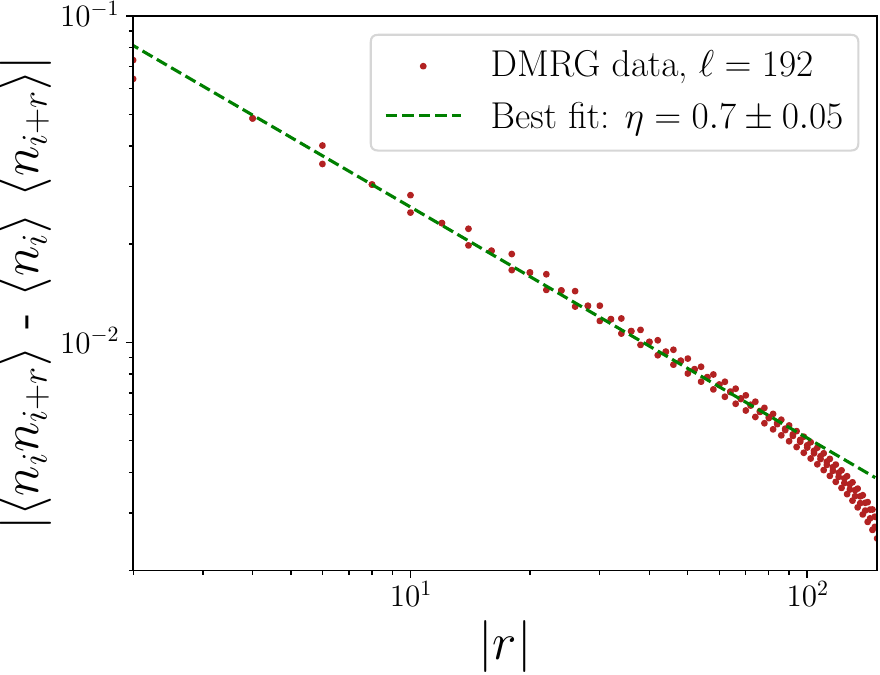}
    \caption{Power law associated with the density-density correlation at $v_{XY}$. $\eta \approx 0.7  \in {1/4, 1}$ is suggestive of proximity to a DQCP (see \cref{subsubsec:numerical_DQCP}). Data is plotted assuming $i = 192$, and $r$ even.}
    \label{fig:DQCP_eta}
\end{figure}

We plot the density-density connected correlator, restricted to the even sites, and find the critical exponent $\eta$ to be $0.7 \pm 0.05$ \cref{fig:DQCP_eta}. This is consistent with $K \approx 0.35 \in (\frac{1}{8}, \frac{1}{2}) $, suggesting deconfined quantum criticality. We caution the reader that this may be an artefact of the system sizes we are dealing with, due to two key reasons: (i) The pinning potentials applied at the edges of the system affect the correlators in the bulk more so in the critical phase, requiring a careful finite size analysis. (ii) It may be that the correlation length of the system is large, but finite, suggesting a weakly-first order transition.

\subsubsection*{XXZ chain limit}
In the limit $V_S \rightarrow + \infty$, even sites are effectively unoccupied $\langle n_i \rangle \ll 1$. Hence, the model reduces to that of hardcore bosons hopping on exclusively odd sites, with the effective hopping parameters, to leading order in perturbation theory, $t' = t^2/\VS$ and nearest neighbor interaction $V' = \VAB - t^2/\VS$. Thus, this provides an alternative way to obtain the perturbative limit suggested in \cref{eqn:tiles_XXZlimit}.

\section{Ground state of \sqQVM{} is described by the tile chain}
\label{appendix:tilechain_groundstatefragment}
Here we show that assuming periodic boundary conditions, the ground state of the tile chain (introduced in \cref{sec:tile_chain}) lies in the sector with no $\tileAS$. We recall the presence of local, conserved quantities 
\begin{equation}
    \Pi_{I, I + 1} = \ket{\tileAS} \bra{\tileAS}_{I, I + 1},
\end{equation}
which allow the energy eigenbasis to be chosen such that each eigenstate has well defined values of $\Pi_{I, I + 1}$, or equivalently, well defined positions of $\tileAS$ tiles.

Suppose, for the sake of contradiction, that $\ket{GS_{n_{AS} > 1}}$ is the ground state of the system, with at least one $\tileAS$. Then, one can construct a corresponding state $\ket{\psi_{n_{AS} = 0 }}$ in the zero $\tileAS$ sector, via the prescription  
\begin{equation}
    \ket{\psi_{n_{AS} = 0}} = \left( \prod_{I} \ket{\hspace{0.004\columnwidth}\tileS}\bra{\tileAS}_{I,I+1} \right) \ket{GS_{n_{AS} > 1}}
\end{equation}
where the product has been taken over $I$ such that $(I, I + 1)$ is the location of $\tileAS$ in $\ket{GS_{n_{AS} > 1}}$. It can be verified that this construction allows one to establish
\begin{equation}
\begin{split}
    \bra{\psi_{n_{AS} = 0}} \hat{K} \ket{\psi_{n_{AS} = 0}} &= \bra{GS_{n_{AS} > 1}} \hat{K} \ket{GS_{n_{AS} > 1}}, \\
    \bra{\psi_{n_{AS} = 0}} \hat{V} \ket{\psi_{n_{AS} = 0}} &= \bra{GS_{n_{AS} > 1}} \hat{V} \ket{GS_{n_{AS} > 1}}.
\end{split}    
\end{equation}
This implies that for all $v$, 
\begin{equation}
\label{eqn:AS_to_S_energy}
    \bra{\psi_{n_{AS} = 0}} \hat{H} \ket{\psi_{n_{AS} = 0}} = \bra{GS_{n_{AS} > 1}} \hat{H} \ket{GS_{n_{AS} > 1}}.
\end{equation}
Suppose the lowest energy state in the sector with zero $\tileAS$ is denoted $\ket{GS_{n_{AS} = 0}}$, then we have
\begin{equation}
\label{eqn:groundstate_bound}
    \bra{GS_{n_{AS} = 0}} \hat{H} \ket{GS_{n_{AS} = 0}} \leq \bra{\psi_{n_{AS} = 0}} \hat{H} \ket{\psi_{n_{AS} = 0}}.
\end{equation}
Thus, \cref{eqn:AS_to_S_energy,eqn:groundstate_bound} establish that if the ground state of the model is unique, then it must lie in the sector without any $\tileAS$. If there do exist degenerate ground states, then at least one of these must belong to the sector without any $\tileAS$. Thus, to investigate the ground state properties of the model for various values of $v$, it is sufficient to restrict oneself to the sector with no $\tileAS$, which in turn maps to the tile model with $v_{AB} = v_{S} = \sqrt{2}v$. 

\section{Exact diagonalization for the tile chain}
\begin{figure}
    \centering
    \includegraphics[width=\columnwidth]{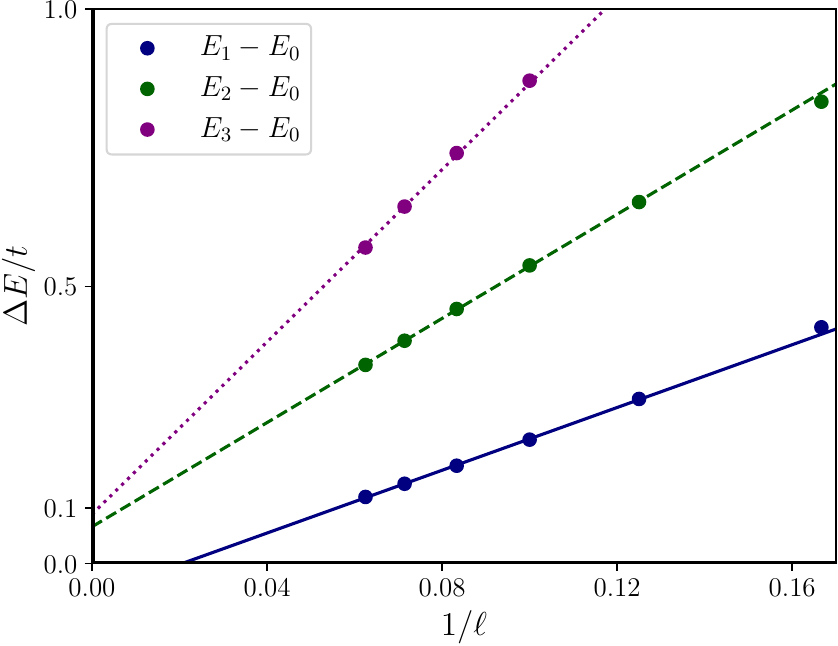}
    \caption{Finite-size scaling analysis of the gap in the tile model for parameters $\VAB = \VS = 0$. $E_0, E_1, E_2, E_3$ are the four lowest energy eigenvalues (obtained via exact diagonalization) for a chain of length $\ell$ with periodic boundary conditions.}
    \label{fig:tile_gap_00}
\end{figure}

We exactly diagonalize the tile hamiltonian with periodic boundary conditions for parameters $\VAB = \VS = 0$, and for system sizes $\ell = 6, 8, 10, 12, 14, 16$. We find that finite-size scaling (see \cref{fig:tile_gap_00}) of the first few excitation energies are consistent with a gapped, doubly degenerate, VBS phase. The gap is found to be of the order $\Delta E \sim 0.1 t$.

\section{Comparison of ground state energies for all four models}
\begin{figure}
    \centering
    \includegraphics[width=\columnwidth]{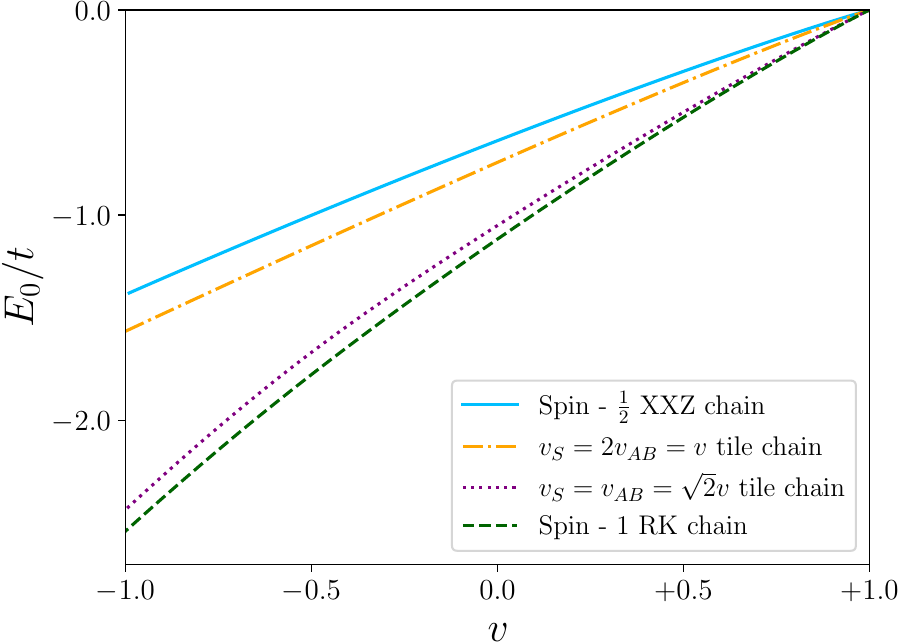}
    \caption{Comparison of ground state energies for the four models of interest.}
    \label{fig:energy_comparison}
\end{figure}

Several of the quasi-1D RK models we consider map to different spin-chains in different sectors. This opens up the possibility of a first-order transition, due to a level crossing between the ground state energies of different sectors of the model. To rule out this possibility when $v < 1$, we numerically determine the ground state energy density for the spin-1 RK chain, and the tile chain for $v_{AB} = v_{S} = \sqrt{2} v$ and $v_{AB} = v_{S} = \sqrt{2} v$. We use the exact expression for the ground state energy of the XXZ model \cite{XXZ_YangYang1,XXZ_YangYang2} to plot \cref{fig:energy_comparison}. We find that there are no level-crossings for $v < 1$, but that all spin-chains become degenerate with $0$ energy at the RK point. This is consistent with the large ground state degeneracy expected for $v \geq 1$, and the first-order transition at $v = 1$.

\label{appendix:energy_comparison}

\section{Spin-1 RK chain as sum of projectors}
\label{appendix:spinone_sumofprojectors}
At the RK point, the spin-1 RK hamiltonian can be written as a sum of projectors,
\begin{align}
    H \propto \sum_{I} \ket{\psi_1}\bra{\psi_1} + \ket{\psi_2}\bra{\psi_2} + \ket{\psi_3}\bra{\psi_3} + \ket{\psi_4}\bra{\psi_4}.
\end{align}
where $\ket{\psi_i}$ are given by
\begin{align*}
    \sqrt{2} \ket{\psi_1} &= \ket{\uparrow ~ 0} - \ket{0 ~ \uparrow}, \\
    \sqrt{2} \ket{\psi_2} &= \ket{\downarrow ~ 0 } - \ket{0 ~ \downarrow}, \\
    \sqrt{2} \ket{\psi_3} &= \ket{\uparrow ~ \downarrow} - \ket{0 ~ 0}, \\
    \sqrt{2} \ket{\psi_4} &= \ket{\downarrow ~ \uparrow} - \ket{0 ~ 0}.
\end{align*}

From the above one can verify that the states from Eq.\eqref{RK-zstates} are exact eigenstates for any $z$.

\section{Details of local Hilbert space fragmentation}
\label{appendix:hilbertspacefragmentation}

\subsection{Two-string \sqQVM{}}
In \cref{subsec:tilechain_sqQVM}, we show that in the $n_x = 2$ sector, configurations of the quantum six vertex model compactified onto a narrow torus (\cref{fig:TT_QVM}) have an explicit correspondence (\cref{fig:sqQVM_dict}) to tilings generated by $\tileA, \tileB, \tileX, \tileY$. The Hilbert space dimension of the model grows exponentially with system size $|\mathcal{H}| \sim a d^{\ell}$, where $a$ depends on boundary conditions imposed on the model. $d$ can be determined by counting how the number of tilings $c_\ell$ of a strip of length $\ell$ grows. $c_\ell$ can be shown to satisfy the recursion relation
\begin{equation}
    c_{\ell + 2} = 2 ~ c_{\ell + 1} + 2 ~ c_{\ell},
\end{equation}
which can be explicitly solved, and yields
\begin{equation}
    c_{\ell} = \frac{\left( 1 + \sqrt{3} \right)^{\ell + 1} - \left( 1 - \sqrt{3} \right)^{\ell + 1} } {2 \sqrt{3}}.
\end{equation}
Thus, we have $d = 1 + \sqrt{3} \approx 2.73$.

\subsection{Tile chain}
The tile chain introduced in \cref{sec:tile_chain} possesses a Hilbert space whose dimension scales exponentially with system size $|\mathcal{H}| \sim a d^{\ell}$, where $a$ depends on boundary conditions imposed on the model. $d$ can be determined by counting how the number of tilings $c_\ell$ of a strip of length $\ell$ grows. $c_\ell$ can be shown to satisfy the recursion relation
\begin{equation}
    c_{\ell + 2} = 2 ~ c_{\ell + 1} + 1 ~ c_{\ell},
\end{equation}
which can be explicitly solved, and yields
\begin{equation}
    c_{\ell} = \frac{\left( 1 + \sqrt{2} \right)^{\ell + 1} - \left( 1 - \sqrt{2} \right)^{\ell + 1} } {2 \sqrt{2}}.
\end{equation}
Thus, we have $d = 1 + \sqrt{2} \approx 2.41$.

\subsection{Number of Hilbert space fragments}
For the model introduced in \cref{subsec:tilechain_sqQVM}, the projector $\Pi_i$ acting on positions $i$ and $i + 1$ as $\ket{\tileAS} \bra{\tileAS}$ commutes with the hamiltonian, which leads to multiple disconnected sectors in the theory. While the operators $\Pi_i$ commute with each other, they are not independent and are constrained by the relation $\Pi_i \Pi_{i + 1} = 0$. As a consequence, the number of sectors labeled by distinct eigenvalues of $\Pi_i$, grow as $\sim a d^{\ell}$, where $a$ depends on the boundary conditions imposed on the model. $d$ can be calculated by explicitly counting the number of ways $\tileAS$ can be placed on a strip of length $\ell$, denoted $c_\ell$, which satisfies the recursion relation
\begin{equation}
    c_{\ell + 2} = c_{\ell + 1} + c_{\ell}.
\end{equation}
This recursive relations, along with the conditions $c_{1} = 1, c_{2} = 2$, define the Fibonacci numbers $c_{\ell} = F_{\ell}$, which are known to scale as $\varphi^\ell$. Thus, $d$ is the golden-ratio 
\begin{equation}
    d = \varphi = \frac{- 1 + \sqrt{5}}{2} \approx 1.62.
\end{equation}

\Appendix{Path integral study of the RK point in the Spin - 1 model}\label{appendix:path_integral}

Define 
\begin{equation}
    \ket{z} = \frac{\ket{\uparrow} + z\ket{0} + z^2 \ket{\downarrow}}{\sqrt{1 + |z|^2 + |z|^4}}.
\end{equation}
Suppose, there exists a positive measure $\mu$ such that
\begin{equation}
    \int \frac{dz^* dz}{2 \pi \ci} \mu(z) \ket{z} \bra{z} = \mathbbm{1}.
\end{equation}
To write down an imaginary time path integral, we insert an To write down an imaginary time path integral for the partition function of this quantum problem, we require the following results

\begin{equation}
\begin{split}
    &\braket{{\mathbf{z}(\tau + \Delta \tau) | \mathbf{z}(\tau)}} = \\
    &  1 + \Delta \tau \sum_i \frac{1 + 2 |z_i|^2}{1 + |z_i|^2 + |z_i|^4} \frac{\partial_\tau z^*_i z_i - z^*_i \partial_\tau z_i }{2} + \mathcal{O}(\Delta \tau^2) 
\end{split}
\end{equation}
where $\tau$ dependence has been suppressed on the right hand side. For the RK point $v = 1$, we have

\begin{equation}
\begin{split}
    &\braket{{\mathbf{z}(\tau + \Delta \tau) | e^{- \Delta \tau \ket{H}} | \mathbf{z}(\tau)}} = \braket{{\mathbf{z}(\tau + \Delta \tau) | \mathbf{z}(\tau)}} \\
    &\times \left( 1 - \Delta \tau ~ t \sum_i \frac{(1 + |z_i|^2)(1 + |z_{i+1}|^2}{1 + |z_i|^2 + |z_i|^4} (z_i - z_{i +1})^2 + \ldots \right)
\end{split}
\end{equation}
where we have neglected terms of the order $O(z_i - z_{i+1})^3$, as we are interested in the long-wavelength low energy theory. Thus, the partition function for the problem at temperature $\beta$ can be written as $\mathcal{Z} = \int \mathcal{D}\ [\mathbf{z^*}, \mathbf{z}] e^{-S[\mathbf{z^*},\mathbf{z}]}$ where the action reads

\begin{equation}
\label{eqn:action}
\begin{split}
    S[\mathbf{z^*},\mathbf{z}] = \int_0^\beta d \tau \int_0^{\ell} dx ~ & \frac{1 + 2 |z_i|^2}{1 + |z_i|^2 + |z_i|^4} \frac{\partial_\tau z^*_i z_i - z^*_i \partial_\tau z_i }{2} \\
    +& t  \frac{(1 + |z_i|^2)(1 + |z_{i+1}|^2)}{1 + |z_i|^2 + |z_i|^4} (\partial_x z)^2
\end{split}
\end{equation}

We know that $z(x, \tau) = z_0$ (constant) is a saddle point (In fact, it is the vacuum of the theory, by construction of the RK point!). We study the excitations about this state by expanding about the saddle point $z(x, \tau) = z_0 + \zeta(x,t)$ where $|\zeta(x,t)| \ll |z_0|, 1$. The first term in \cref{eqn:action} corresponds to a berry phase, and the corresponding berry curvature can be calculated
\begin{equation}
    \mathcal{B}(z_i) = 2 \frac{1 + 4 |z_i|^2 + |z_i|^4}{1 + |z_i|^2 + |z_i|^4}
\end{equation}
Which allows the first term to be rewritten (after subtracting a total derivative in time) as $\mathcal{B}(z_0) (\zeta^* \partial_\tau \zeta - \partial_\tau \zeta^* \zeta)/4$. Thus, up to quadratic order in $\zeta$, the action can be written as

\begin{equation}
\label{eqn:new_action}
\begin{split}
    S[\mathbf{\zeta^*},\mathbf{\zeta}] &= \int_{S^1 \times S^1} d\tau dx ~ \frac{1 + 4 |z_0|^2 + |z_0|^4}{1 + |z_0|^2 + |z_0|^4}  \\
    & \times \left( \frac{\partial_\tau \zeta^*_i \zeta_i - \zeta^*_i \partial_\tau \zeta_i }{2} + t  \frac{(1 + |z_0|^2)^2}{1 + 4|z_0|^2 + |z_0|^4} |\partial_x \zeta|^2 \right)
\end{split}
\end{equation}

Thus, the low energy dispersion ($k \ll 1)$ is given by

\begin{equation}
    \omega = t \frac{(1 + |z_0|^2)^2}{1 + 4 |z_0|^2 + |z_0|^4} k^2
\end{equation}

\begin{figure}[h]
    \centering
    \includegraphics[width=0.8\columnwidth]{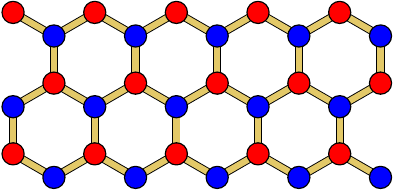}
    \caption{\hxQDM{} which maps to a chain of hardcore bosons with nearest neighbor exclusion. See \cref{appendix:exclusion_chain}}.
    \label{fig:exclusion_chain}.
\end{figure}

\Appendix{Hardcore bosons with nearest neighbor exclusion}
\label{appendix:exclusion_chain}
The \hxQDM{} shown in \cref{fig:exclusion_chain}, is compactified onto a narrow torus of length $\ell_x$, and possesses $2 \ell_x$ plaquettes. The model maps to a hardcore boson model with the following hamiltonian:
\begin{equation}
\begin{split}
    \ham    =& -t \sum_{i = 1}^{\ell_x} b_{i + 1}^\dagger b_i + h.c. \\
            & + V \sum_{i = 1}^{\ell_x} (n_i - n_{i+2})^2 \\
            & + \infty \sum_{i = 1}^{\ell_x} n_i n_{i + 1}.
\end{split}
\end{equation}
Since boson number is a conserved quantity under this hamiltonian, the operator can be resolved into the respective sectors, and under which we find the correspondence:
\begin{equation}
    \ham (n, \ell) \equiv H_{XXZ} (n, \ell - n).
\end{equation}

So in some sense, each sector of the hamiltonian is the same as that of a shortened XXZ chain. However the correlators and the effective field theory description of the model will be distinct. For more details, see \cite{constrainedXXZ} or \cite{verresenVishwanathPollmann}.

\section{Alternate description: \hxQDM, single non interacting string}
\label{subsec:XXZ_hxQDM}

\begin{figure}
    \centering
    \includegraphics[width = 0.9\columnwidth]{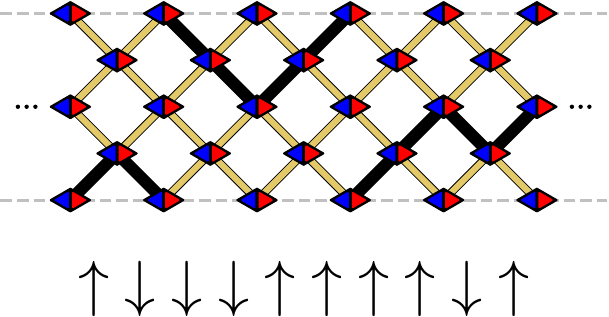}
    \caption{Alternative way to map model discussed in \cref{subsec:hxQDM_single_string} to an XXZ chain.}
    \label{fig:XXZalternate}
\end{figure}

The \hxQDM{} shown in \cref{fig:XXZ_collection}(c), is compactified onto a narrow torus of length $\ell_x$, and possesses $2 \ell_x$ plaquettes. Instead of grouping vertices along the links oriented up-right (as done in \cref{subsec:hxQDM_single_string}), one can group form composite vertices along the horizontal links to get a string description on the square lattice \cref{fig:XXZalternate}.

The sector $n_x = 0$ is inert and has a unique state, while the sector $n_x = 2$ has a degeneracy of $2^{\ell_x}$. The $n_x = 1$ sector maps  onto an XXZ chain on $\ell_x$ sites, by identifying segments as shown in \cref{fig:XXZalternate}, with an additional specification of the $y$-momentum sector $k_y \in \{0, \pi\}$, manifesting as an internal flux in the XXZ ring. Thus the $1 + 3 \cdot 2^{\ell_x}$ dimensional Hilbert space of the problem reduces to
\begin{equation}
    \mathcal{H} = \bigoplus \begin{cases}
                    \mathbf{0}^{} & n_x = 0 \\[0.2em]
                    \bigoplus_{k_y} \XXZ{\ell_x}{k_y} & n_x = 1 \\[0.2em]
                    \mathbf{0}^{\otimes 2 ^{\ell_x}} & n_x = 2.
                    \end{cases}
\end{equation}

\clearpage
\bibliography{RK_models}
\end{document}